\documentclass[apj]{emulateapj}

\usepackage[T1]{fontenc}
\usepackage{natbib}
\usepackage{epstopdf}
\usepackage{epsfig}
\usepackage{txfonts}
\usepackage{mathrsfs}
\usepackage{natbib,caption2}
\usepackage{subfigure}
\usepackage{longtable}
\usepackage{graphicx}
\usepackage{graphics}
\usepackage{keyval}
\usepackage{trig}
\usepackage{float}
\usepackage{url}
\usepackage{hyperref}
\usepackage[dvipsnames]{color}
\usepackage{lineno}
\usepackage{threeparttable}

\submitted{Accepted for publication in {\it The Astrophysical Journal}, 2023 October 28}

\newcommand \beq{\begin{equation}}
\newcommand \eeq{\end{equation}}
\newcommand \bey{\begin{eqnarray}}
\newcommand \eey{\end{eqnarray}}

\newcommand \tide{\, {\rm r_{\rm t} } }

\newcommand \re{\, {\rm R_{\rm h}} }

\newcommand \kpc{\, {\rm kpc} }
\newcommand \Mpc{\, {\rm Mpc} }
\newcommand \msun{{\rm M}_\odot}

\newcommand \kms{\, {\rm km \, s}^{-1} }

\newcommand \ar{a_-/a_+}
\newcommand \rvir{R_{\rm vir}}

\begin{document}
\title[Lopsidedness and LRA in simulated galaxies]{Shape asymmetries and the relation between lopsidedness and radial alignment in simulated galaxies}
\author{Jinzhi Shen\altaffilmark{1,2}}
\author{Xufen Wu\altaffilmark{1,2}}
\thanks{Email: xufenwu@ustc.edu.cn}
\author{Yirui Zheng\altaffilmark{3}}
\author{Beibei Guo\altaffilmark{1,2}}
\altaffiltext{1}{CAS Key Laboratory for Research in Galaxies and Cosmology, Department of Astronomy, University of Science and Technology of China, Hefei 230026, China; }
\altaffiltext{2}{School of Astronomy and Space Science, University of Science and Technology of China, Hefei 230026, China}
\altaffiltext{3}{Department of Astronomy, Shanghai Jiaotong University, Shanghai, P.R. China}

\begin{abstract}
  Galaxies are observed to be lopsided, meaning that they are more massive and more extended along one direction than the opposite. In this work, we provide a statistical analysis of the lopsided morphology of 1780 isolated satellite galaxies generated by TNG50-1 simulation, incorporating the effect of tidal fields from halo centres. The isolated satellites are galaxies without nearby substructures whose mass is over $1\%$ of the satellites within their virial radii. We study the radial alignment (RA) between the major axes of satellites and the radial direction of their halo centres in radial ranges of $0$-$2\re$, $2$-$5\re$ and $5$-$10\re$ with $\re$ being the stellar half mass radius. According to our results, the RA is virtually undetectable in inner and intermediate regions, yet it is significantly evident in outer regions. We also calculate the far-to-near-side semi-axial ratios of the major axes, denoted by $\ar$, which measures the semi-axial ratios of the major axes in the hemispheres between backwards (far side) and facing (near side) the halo centres. In all the radial bins of the satellites, the numbers of satellites with longer semi-axes on the far side are found to be almost equal to those with longer semi-axes on the near side. Therefore, the tidal fields from halo centres play a minor role in the generation of lopsided satellites. The long semi-major-axes radial alignment (LRA), i.e., an alignment between the long semi-major-axes of satellite galaxies and the radial directions to their halo centres, is further studied. No clear evidence of LRA is found in our sample within the framework of $\Lambda$CDM Newtonian dynamics. Finally, we briefly discuss the possible origins of the asymmetry of galaxies in TNG50-1.

\end{abstract}

\section{Introduction}

The mass distributions of gaseous and stellar components in many galaxies are observed to be lopsided. Since the 1960s, optical or HI observations have revealed that some spiral galaxies are asymmetric between their northern and southern sides \citep[e.g.,][]{Dieter1962,Arp1966,Beale_Davies1969,Rogstad1971,Bosma1978,Sancisi_Allen1979}. \citet{Baldwin+1980} were the first to find that in a sample of about twenty lopsided galaxies, most of them are isolated galaxies. Later, systematic studies on the observed disc galaxies showed that about $1/3$ of field disc galaxies are significantly lopsided with an azimuthal $m=1$ Fourier amplitude $A_1 \ge 0.20$ for the stellar light \citep{Rix_Zaritsky1995,Zaritsky_Rix1997}. In a sample of 54 early-type disc galaxies (S0 to Sab), $20\%$ of them are lopsided, with $A_1 \ge 0.19$ for the R-band surface brightness \citep{Rudnick_Rix1998}. \citet{Conselice+2000} and \citet{Bournaud+2005b} found that the lopsidedness of galaxies strongly correlates with the morphological type, that late-type discs and irregulars are more asymmetric and ellipticals and lenticular galaxies are less lopsided. However, \citet{Angiras+2006} unveiled that the early-type galaxies are more lopsided than the late-type galaxies. Moreover, the lopsidedness of galaxies strongly correlates with stellar surface density and tends to increase with radius, as revealed by a systematical study on the light distribution of 25,155 present-day galaxies from the Sloan Digital Sky Survey Data Release 4 \citep[SDSS DR4,][]{Reichard+2008}. It is also confirmed that the lopsidedness of stellar light distribution corresponds to the lopsided stellar mass distribution \citep{Reichard+2008}. About $50\%$ in the Westerbork HI sample of Spiral and Irregular Galaxies (WHSIP) survey are observed to be strongly lopsided within small radii, $R_{\rm 25}$, \citep[i.e., 2.5 times of disc scale lengths,][]{vanEymeren+2011}, and there are $20\%$ galaxies in this sample displaying noticeably increasing lopsidedness out to large radii. In a most recent study on the HI emission line of 29,958 nearby galaxies by \citet{Yu+2022}, $20\%$ of the sample of galaxies are significantly asymmetric considering systematic bias due to S/N. Overall, a certain fraction of galaxies are morphologically lopsided in observations.

The lopsided morphologies of galaxies are presumably caused by a time-dependent non-equilibrium dynamical state. Several mechanisms have been proposed to understand the morphological lopsidedness of galaxies within the framework of standard Newtonian dynamics. Such mechanisms are divided into two classes, namely internal and external ones \citep[eg. a review of lopsided galaxies, ][]{Jog_Combes2009}. The internal mechanisms include the self-gravitational instability \citep{Zaritsky+2013}, the offset centres between the disc and the dark matter halo in a galaxy \citep{Levine_Sparke1998,Noordermeer+2001}, and the lopsided dark matter host halo that provides a lopsided internal potential \citep{Weinberg1994,Jog1997}.

The external mechanisms include galaxy mergers \citep{Walker+1996,Zaritsky_Rix1997,Bok+2019}, asymmetric gas accretion from cosmological filaments \citep{Bournaud+2005b,Mapelli+2008}, close tidal encounters \citep{Kornreich+2002} and ram pressure stripping \citep{Mapelli+2008,Scott+2010,Kenney+2015}. \citet{Bournaud+2005a} showed that the lopsidedness induced by minor mergers disappears when the companion is disrupted and most of the lopsided galaxies are not undergoing mergers. In a flyby interaction between a disc target galaxy and a companion galaxy \citep{Mapelli+2008}, the lopsided feature can be produced in the target galaxy. Since the disc galaxy is rotating, the perturbation appears on the opposite side of the intruder galaxy after half the rotating period (approximately 300 Myr) and this explains well the observed lopsided galaxy NGC 891. Moreover, \citet{Mapelli+2008} indicated that the overall stellar component is not lopsided in the case of gas accretion, and that ram pressure only generates moderate tidal tails. The observational results that a larger fraction of lopsided galaxies are early-type implies a tidal origin for the lopsidedness \citep{Angiras+2006,vanEymeren+2011}. However, the lopsidedness does not correlate with the tidal parameter, $T_p$ \citep{vanEymeren+2011}, which quantifies the effect of tidal force.

Within the framework of $\Lambda$CDM Newtonian dynamics, galaxies are supposed to be embedded in dark matter haloes. \citet{Ciotti_Dutta1994} made the first prediction that the tidal force from a host halo causes an alignment between the major axes of satellites and the radial direction of the central dark matter haloes. As satellites orbit around the gravitational centre of their host halos, they are continuously reshaped by tidal fields. The tidal forces elongate the mass distribution of a satellite along the radial direction and squeeze the satellite along the directions perpendicular to the radial direction. This prediction appears to agree with the early observations that the projected major axes of galaxies preferentially align with the radial direction of the cluster centres \citep{Hawley_Peebles1975,Thompson1976}. The radial alignment (hereafter short for RA) between galaxies and cluster centres has been further confirmed in the SDSS observations \citep{Pereira_Kuhn2005,Faltenbacher+2007a,Wang+2008} and in the cosmological simulations \citep{Faltenbacher+2008,Pereira+2008,Knebe+2020}. However, the transit through the pericenter of the orbit of the satellite is so rapid that the time could not be long enough for the satellite galaxy to realign itself towards the host halo centre. Indeed, some other observations \citep{Hung_Ebeling2012,Schneider+2013,Chisari+2014,Sifon+2015} point towards the absence of RA in large samples of clusters at different redshifts. More recently, \citet{Singh+2015} claimed to find an RA in a sample of early-type galaxies in the SDSS-III. \citet{Wang+2019} also observed a strong signal for radial alignment in satellites orbiting host galaxy pairs in the SDSS 13. Since the lopsidedness and the RA can both be explained by the tidal force from a massive gravitational source, a natural question arises as to whether the long semi-major axis of a lopsided galaxy aligns with the radial direction of the cluster centre. Tidal forces exerted by a halo centre are roughly symmetrical along both near and far sides of a satellite relative to its halo centre. Nevertheless, when considering the higher-order terms of the Taylor expansion for the gravitational force produced by the halo, an asymmetry in these forces emerges. This asymmetry could potentially result in a satellite galaxy adopting an uneven shape.

In principle, the dark matter halo provides a deep gravitational potential. The lopsided shapes and the RA of satellite galaxies indicate that the semi-axial ratio of the major axis $\ar < 1.0$, which means that the semi-axis facing to the halo centre is longer than that on the opposite side. However, the asymmetry in the shape of a satellite is a subtle consequence of the higher-order terms in gravitational calculations. It remains uncertain whether this asymmetry can be consistently observed in cosmological simulations. The accurate definition of $a_-$ and $a_+$ are to be presented in \S \ref{semiaxial}. However, these problems have not yet been systematically addressed in the framework of $\Lambda$CDM cosmological simulations. The existing studies on the RA only focus on the alignment angles between the overall major axes and the radial direction to the cluster centres. We shall provide a more in-depth analysis below.

In this work, we present a detailed analysis of the asymmetric shapes of present-day satellite galaxies from the public data release of the cosmologically simulated galaxies, TNG50-1 \citep{Nelson+2019}. We aim to examine whether LRA exists in these satellites. The manuscript is organised as follows. In \S \ref{models}, a sample of isolated satellite galaxies is selected. The major axes of the satellites are determined by calculating the eigenvectors of the inertia tensor. Then the lopsidedness of satellites is analysed based on the semi-major axial ratios $a_-/a_+$ \S \ref{axialratio}. The existence of RA and LRA are investigated in \S \ref{alignment}. Furthermore, a table of strongly lopsided satellites in the TNG50-1 simulation is provided and the mechanisms to generate the lopsidedness are discussed in \S \ref{discussions}. Finally, our results are summarised in \S \ref{conclusions}.

\section{Simulated satellite galaxies and their axial ratios}\label{models}
The present-day ``galaxies'' studied here are obtained from the TNG50-1 simulation, which is part of the IllustrisTNG simulation suite \citep{Marinacci+2018,Naiman+2018,Nelson+2018,Nelson+2019,Springel+2018,Pillepich+2018}. The IllustrisTNG project, using a moving-mesh code {\it AREPO} \citep{Springel+2010}, is the next generation of the magnetohydrodynamical (MHD) cosmological simulation, Illustris \citep{Vogelsberger+2014b,Vogelsberger+2014a}. In the TNG suite of cosmological simulations, the TNG50-1 is recognized for having the highest resolution within a box size of $51.7 \Mpc$ \citep{Nelson+2019}. There are $2160^3$ gas particles and also $2160^3$ dark matter particles in the TNG50-1 simulation, and the mass resolutions for gas and dark matter are $8.5\times 10^4\msun$ and $4.5\times 10^5\msun$, respectively. Such a high resolution allows us to extract satellite galaxies with high enough resolution to analyse their shapes.

\subsection{Sample selection}\label{selection}
We select the satellite galaxies at z=0 using the following criteria. First of all, the haloes at the boundary of the simulation box, i.e., if there are stellar particles in a halo that travel through the periodic boundary and appear at the other side of the box, are excluded in our halo sample. The satellite galaxies are then selected from the haloes containing at least one satellite galaxy. Note that the central galaxies in the haloes are not included in our sample. The satellites studied in this work are restricted within a stellar mass range of $[10^8, ~10^{11}] \msun /h$. We carefully checked the number of stellar particles in each galaxy model. Even a satellite galaxy with a mass of $10^8\msun$ contains at least $2400$ stellar particles, which ensures the shapes of galaxies calculated from the stellar particles are reliable. Moreover, the radial distances between the satellites and their halo centres are further than $0.3\rvir$. Here $\rvir$ is the virial radius of a halo, within which the mean overdensity of the halo mass is $200$ times the critical density of the Universe, i.e., $\rho_{\rm crit}=\frac{3H_0^2}{8\pi G}$ and $H_0=67.74 \kms {\rm Mpc}^{-1}$ in the TNG50-1 simulation. There are $2923$ galaxies that satisfy the above selection constraints.

Finally, to overcome the perturbation from recent mergers or collisions, we require the satellite galaxies to be isolated, which means that there are no nearby galaxies or substructures with a mass over $1\%$ of a satellite within the virial radius of the subhalo, $r_{\rm vir,sub}$. The final sample contains $1780$ isolated galaxies. The isolated satellites do not refer to field galaxies, but satellites without any nearby galaxies within $r_{\rm vir,sub}$ of themselves. $r_{\rm vir,sub}$ is the truncation radius for the subhalo of a satellite, within which the mean overdensity of the subhalo mass is $200$ times the critical density of the Universe in the TNG50-1 simulation. Apart from the isolated galaxies, in this work, we define the rest of $1409$ selected satellites with nearby substructures from the same halo within their $r_{\rm vir,sub}$ as non-isolated galaxies.

We shall study the semi-axial ratio of the major axes and then analyse the LRA of the satellite galaxies in the following sections. We shall also provide a comparison between the isolated and non-isolated samples of satellites.

\subsection{Intrinsic major axes of the satellites}\label{pa}
To study the lopsidedness of satellite galaxies, we need to find out the intrinsic major axes. The centre of a satellite is defined by the densest region of the stellar component. The principle axes of a satellite are determined in a radial range of $0$-$2\re $ for the inner regions, $2$-$5\re $ for the intermediate regions and $5$-$10\re $ for the outer regions by the following approach. Here $\re$ is the half-mass radius of the stellar component in a galaxy. Since there are fewer particles in the larger radius of a galaxy, the particle noise increases as the radius grows. In the intermediate and outer radial bins, we only calculate the principle axes of a galaxy if there are more than a thousand particles in the corresponding radial bins. In the beginning, we assume that the galaxy is a triaxially symmetric system whose isodensity surface can be described by the ellipsoidal equation, $\left(\frac{x}{a}\right)^2 + \left(\frac{y}{b}\right)^2 +\left(\frac{z}{c}\right)^2=r^2$. Initially, $(x,~y,~z)$ are the coordinates of stellar particles of a satellite obtained from the TNG50-1 simulation, and $a,~b$ and $c$ are the characteristic scale lengths and take the value of unity. $r$ is the distance to the satellite centre.

The moments of inertia tensor of stellar particles inside a sphere with a radius of $r$ are calculated by
\bey
I_{\rm xx} (r) &= &\frac{\sum_i m_i(y_i^2+z_i^2)}{\sum_i m_i}\\
I_{\rm xy}(r) &= &\frac{\sum_i -m_ix_iy_i}{\sum_i m_i}
\eey
and similar expressions for other components. The diagonalised inertia tensor is computed, and the eigenvalues of the system, $I_{\rm x'x'},~I_{\rm y'y'}$ and $I_{\rm z'z'}$, are obtained. The principle axes of the satellite align with the eigenframe. The characteristic scale lengths are now updated to
\bey
a'(r) =\sqrt{[I_{\rm y'y'}+I_{\rm z'z'}-I_{\rm x'x'}]/2},\\ 
b'(r) =\sqrt{[I_{\rm x'x'}+I_{\rm z'z'}-I_{\rm y'y'}]/2},\\ 
c'(r) =\sqrt{[I_{\rm x'x'}+I_{\rm y'y'}-I_{\rm z'z'}]/2}.
\eey
The intrinsic major, intermediate and minor axes are the $x'-$, $y'-$ and $z'-$axes in the eigenframe of the inertia tensor, and the values of $a'(r),~b'(r)$ and $c'(r)$ are the corresponding intrinsic characteristic scale lengths, respectively.

\subsection{Semi-axial ratios of the satellites}\label{semiaxial}

For a system that has rotated into its eigenframe of the inertia tensor, the positive $x'$-axis is defined as the major axis pointing to the halo centre. Thus the far- and near-side hemispheres are the hemispheres backward and facing the cluster centre, respectively, segmented by a plane perpendicular to the major axis at the centre of the satellite. The characteristic scale lengths of the major axis on the far- and near-sides are described by the root mean square semi-axes, $a_-$ and $a_+$, which are
\bey
a_-(r)=\left(\frac{\sum_i m_ix_i'^2}{\sum_i m_i}\right)^{1/2},~~~~x_i'<0,\\
a_+(r)=\left(\frac{\sum_i m_ix_i'^2}{\sum_i m_i}\right)^{1/2},~~~~x_i'>0,
\eey
respectively. For a perfectly triaxially symmetric system, $a_-$ and $a_+$ should be equal. However, as aforementioned, the galaxies are lopsided, i.e., $a_-$ and $a_+$ can be different. Here we use the semi-axial ratio, $\ar$, to quantify the lopsidedness of a satellite galaxy. The semi-axial ratios are calculated within the radial bins of $0$-$2\re$, $2$-$5\re$ and $5$-$10\re$ for the sample of isolated galaxies.

In addition, in the whole sample of satellite galaxies, the non-isolated galaxies are strongly influenced by the tidal fields from the nearby massive galaxies, i.e., locating within $10\re$ of the satellite galaxy. For non-isolated galaxies, the tidal fields come from both the halo centres and the nearby galaxies. Thus the definition of positive $x'$-axis is different from that of the isolated galaxies. For a non-isolated satellite, the near side is defined by the direction of the object, be it a nearby massive galaxy or the halo centre, along which the tidal field is stronger. To make a comparison to the isolated sample, the semi-axial ratios of the non-isolated galaxies are also computed within the radial bins of $0$-$2\re$, $2$-$5\re$ and $5$-$10\re$.

\begin{table*}\centering
\caption[]{ The statistics of semi-axial ratios of major axes, $\ar$, of the satellites within the radial bins of $0$-$2\re$, $2$-$5\re$ and $5$-$10\re$ in mass bins of $10^8$-$10^9\msun /h$, $10^9$-$10^{10}\msun /h$ and $10^{10}$-$10^{11}\msun /h$ for the isolated ($3^{\rm rd}$-$7^{\rm th}$ columns) and non-isolated ($8^{\rm th}$-$12^{\rm th}$ columns) samples. Here $N_-$ and $N_+$ denote the numbers of galaxies with semi-axial ratios $\ar <1$ and $\ar >1$, respectively. The mean values, $\mu$, and the standard deviation, $\sigma$, for the semi-axial ratios, together with the skewness, are listed in the table. The last column displays the p-value of the Anderson-Darling 2-sample test for the isolated and non-isolated galaxies.
}
\begin{tabular}{lcccccccccccc} 
   \hline & & \multicolumn{5}{c}{Samples: isolated}& \multicolumn{5}{c}{Non-isolated} & AD 2-sample \\ 
  \hline
   Mass  & Radial range & $N_{-}$ & $N_{+}$ & $\mu$ & $\sigma$ & Skewness  & $N_{-}$ & $N_{+}$ & $\mu$ & $\sigma$ & Skewness & p-value \\
   $(M_{\odot}/h)$ & ($R_{h}$) & & & & & & & \\
   \hline $10^8$-$10^9$ & 0-2 & 625 & 636 & 1.00 & 0.05&-0.03  & 475 & 462 & 1.00 & 0.12 & 3.50 & 0.0005 \\
  & 2-5 & 384 & 359 & 1.00 & 0.05 & 0.45  & 309 & 252 & 1.00 & 0.12 & 11.07 &0.0040 \\
  & 5-10 & 56 & 54 & 1.00 & 0.05 & -0.01 & 27 & 21 & 0.99 & 0.07 & -0.33 &0.3295\\
  \hline $10^9$-$10^{10}$ & 0-2 & 211 & 197 & 1.00 & 0.06 & 2.37 & 186 & 173 & 1.00 & 0.08& 0.44 &0.0445\\
  & 2-5 & 192 & 216 & 1.00 & 0.05 & 0.24 & 184 & 174 & 1.00 & 0.06 & 0.34 & 0.1095\\
  & 5-10 & 121 & 120 &1.00 & 0.06 & 0.76 & 109 & 77 & 0.98 & 0.10 & -0.67 &0.0003\\
  \hline $10^{10}$-$10^{11}$  & 0-2 & 51 & 60 & 1.01 & 0.05 & 1.26 & 64 & 49 & 1.00 & 0.07 & 0.27 & 0.0880\\
  & 2-5 & 54 & 57 & 1.00 & 0.06 & 0.35 & 57 & 56 & 1.00 & 0.06 & -0.08 &0.5655\\
  & 5-10 & 44 & 58 & 1.01 & 0.07 & 0.71  & 63 & 39 & 0.98 & 0.11 & -0.25&0.0055\\
  \hline
\end{tabular}
\label{tab:table_1}
\end{table*}

\begin{figure*}
  \includegraphics[width=170mm]{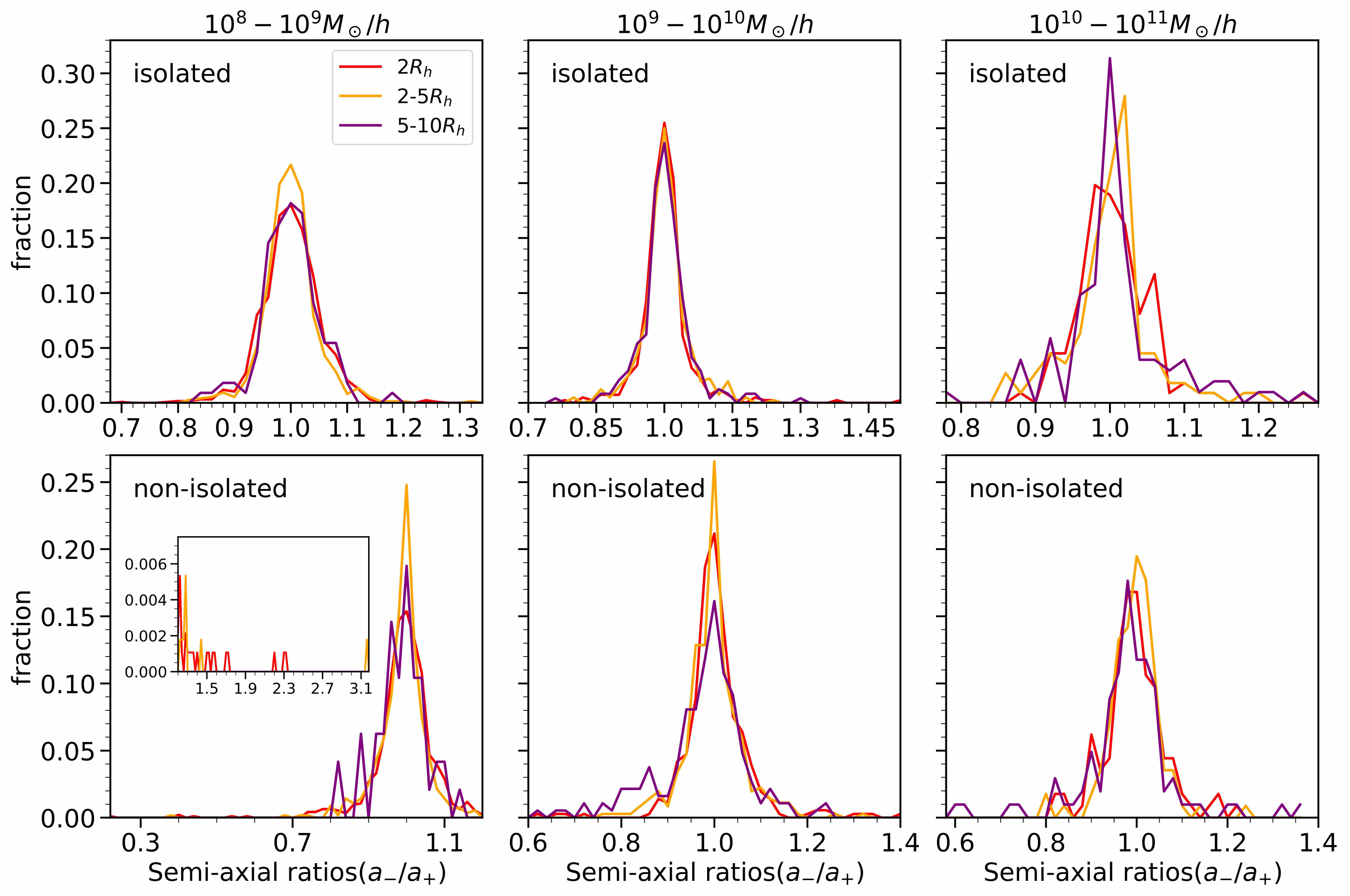}
  \caption{The fraction of satellite galaxies versus their semi-axial ratio, $\ar$, of the major axes in the radial bins of $0$-$2\re$ (red curves), $2$-$5\re$ (golden curves) and $5$-$10\re$ (purple curves) in mass bins of $10^8$-$10^9\msun /h$ (left panels), $10^9$-$10^{10}\msun /h$ (right panels). The upper and lower panels show the results from samples of isolated and non-isolated satellites, respectively.
  }
  \label{gaussianfit}
\end{figure*}

\section{Lopsidedness of the satellites}\label{axialratio}
For a self-gravitationally bound system embedded in a static potential of a dark halo and deviating from the halo centre, the tidal field from the dark halo may change the shape of the self-bound system. The semi-axis facing the halo centre is expected to be longer than the semi-axis backward halo centre. This implies that the system is lopsided and the semi-axial ratio $\ar <1.0$. However, previous studies on tidal effect mainly focused on galaxies interactions such as minor mergers \citep{Walker+1996,Bok+2019} or close encounters \citep{Kornreich+2002}. Considering the tidal force decays with distance as a power law of $D^{-3}$, the tidal effect caused by halo centres is minor. It is unlikely that the tidal field from halo centres can lead to a detectable lopsidedness. However, a systematic test on the lopsidedness of satellite galaxies surrounding their host halos is still lacking. 

Moreover, apart from the tidal interactions due to minor mergers or close encounters, lopsidedness can be produced by several other mechanisms, including such external effects as asymmetric gas accretion \citep{Bournaud+2005b,Mapelli+2008} and ram pressure stripping \citep{Mapelli+2008,Kenney+2015} and such internal effects as self-gravitational instability \citep{Zaritsky+2013}, offset centres between the disc and the dark matter halo \citep{Levine_Sparke1998,Noordermeer+2001}, and the lopsided dark matter host halo \citep{Weinberg1994,Jog1997}.
On the other hand, for a large sample of galaxies, if the tidal fields from the halo centres are not taken into account, the above effects do not produce a preferred alignment for the long semi-major axes with respect to the radial directions of halo centres. The directions of the long semi-major axes are random for a large sample of galaxies. Therefore, the numbers of galaxies with $\ar<1.0$ and $\ar>1.0$ are expected to be approximately equal. However, it remains unclear whether the number of galaxies with long semi-major axes on the near side of halo centres is larger than that on the far side in case the minor effect of the tidal field from the halo centres is incorporated.

An analysis of a large sample of 1912 disc galaxies in Illustris-TNG100 simulation \citep{Lokas2022} has unveiled that only $8\%$ of discs are moderately lopsided with the Fourier amplitude $A_1 \ge 0.10$, and that only $0.37\%$ are strongly lopsided galaxies with $A_1 \ge 0.2$, which is a substantially lower fraction compared to the observations \citep[e.g., $30\% $ in][]{Jog_Combes2009}. In this work, we analyse the lopsidedness of satellite galaxies generated by the TNG50-1 simulation, of which the resolution is higher. It is thus possible to include the low-mass satellites in our sample. Moreover, we study the orientations of the long semi-major axes of the satellites with respect to the radial directions to halo centres (i.e., the LRA), which have not yet been systematically explored.

\subsection{Distribution of the semi-axial ratios}

To examine the distribution of $\ar$ near the value of $1.0$, we calculate $\ar$ of the stellar components of all galaxies in two samples of isolated and non-isolated satellites. Each sample is further divided into three subsamples according to their stellar mass, including low-mass satellites, intermediate-mass satellites and massive satellites with stellar masses being in ranges of $[10^8, 10^{9}] \msun /h$, $[10^9, 10^{10}] \msun /h$ and $[10^{10}, 10^{11}] \msun /h$, respectively. The fractions of galaxies as a function of semi-axial ratios, $\ar$, in bins of $0.02$, of the stellar components are shown in Fig. \ref{gaussianfit}, in the radial bins of $0$-$2\re$ (red curves), $2$-$5\re$ (golden) and $5$-$10\re$ (purple). Moreover, the numbers of galaxies with $\ar <1$ and $\ar >1$, denoted as $N_-$ and $N_+$, are listed in the fourth and fifth columns of Table \ref{tab:table_1} for the samples of isolated and non-isolated satellite galaxies in three mass bins.

For the isolated satellite galaxies (the upper panels in Fig. \ref{gaussianfit}), the fractions of galaxies peaked at around $\ar \approx 1.0$ are almost symmetric. Significant fluctuations appear in the fraction of galaxies versus $\ar$ in the most massive subsample (the upper right panel) since the total number of galaxies within this mass range is the smallest. The almost symmetric shapes of the galaxy fractions indicate that the tidal field from the halo centres plays a minor role in determining the asymmetric shapes of satellite galaxies. For the non-isolated satellites (the lower panels in Fig. \ref{gaussianfit}, the fractions of galaxies also appear to be symmetric around $\ar\approx 1.0$. However, more fluctuations are deviating from $\ar=1.0$. There are more asymmetric satellite galaxies in the non-isolated subsamples. The tidal field effect from a close encounter or a minor merger influences the semi-major axial ratio of a non-isolated satellite more significantly.
Below we introduce a Gaussian distribution function to fit the dependence of fractions of satellites on the varying ratio $\ar$.

\begin{table}\centering
\caption[]{The lopsidedness of satellites is defined by two criteria, namely deviation to the near side semi-major axes $\Delta a$ and the Fourier $m=1$ mode values $A_1$. The fractions of moderately and strongly lopsided satellites are listed in the $2^{nd}-3^{rd}$ and $4^{th} -5^{th}$ columns, respectively.
}
\begin{tabular}{cccccccc}
\hline
 Sample & $f_{\Delta a>0.05}$ & $f_{A_1>0.1}$ &  $f_{\Delta a>0.1}$ & $f_{A_1>0.2}$ \\
\hline
Isolated & $25.9\%$ & $19.9\%$ & $5.8\%$ & $2.5\%$ \\
Non-isolated &$33.6\%$ & $34.2\%$ & $12.8\%$ & $11.1\%$  \\
\hline
\end{tabular}\label{tab:table_2}
\end{table}

We show how the peak of $\ar$ in a Gaussian fitting deviates from 1.0 in Table \ref{tab:table_1}, which quantifies the systematical tidal effect from the halo centres. For the subsample of low-mass galaxies within the mass range of  $[10^8, 10^{9}] \msun /h$, the numbers $N_-$ and $N_+$ are almost equal within all radial ranges, except in the intermediate region, the number of galaxies with $\ar <1.0$ is approximately $10\%$ more than that with $\ar>1.0$. However, the mean values of $\ar$ are $\mu=1.00$ in all ranges of radii, with a standard deviation of $\sigma=0.05$. To judge the shapes of the fractions of galaxies versus semi-axial ratios, we also calculate the skewness of the subsamples,
\beq {\textit Skewness}=\frac{n}{(n-1)(n-2)}\sum_{\rm i=1}^{\rm n}\left[\frac{(\ar)_i-\mu}{\sigma}\right]^3 .\eeq 
We find that the absolute skewness values for this subsample in different radial ranges are all smaller than 0.5, indicating that the distributions of fractions of galaxies are perfectly symmetric around $\ar=1.00$. Thus the long side of the major axes of the low-mass satellites does not point towards the host halo centres. The tidal fields from halo centres do not lead to clearly visible lopsidedness.

In the subsample of low-mass isolated satellites, the tidal fields from halo centres do not lead to clearly visible lopsidedness. Such a conclusion still holds in the subsamples of intermediate-mass and massive isolated satellites. The shapes of the fraction of galaxies versus the semi-axial ratio within the three radial ranges appear to be similar and symmetric. The values of skewness of the fractions of galaxies within $0$-$2\re$ are 2.37 and 1.26 for the intermediate-mass and massive isolated subsamples, respectively, which are positively skewed, with long tails on the right sides. In the upper middle panel of Fig. \ref{gaussianfit}, a few strongly lopsided outliers with semi-axial ratios beyond $5\sigma$ of the subsample lead to the right skewness of fraction of galaxies within $0$-$2\re$. The strongly asymmetric shapes of these galaxies are not caused by the tidal fields from their halo centres. We will discuss the possible mechanisms in \S \ref{origins}.

\begin{figure}
  \includegraphics[width=90mm]{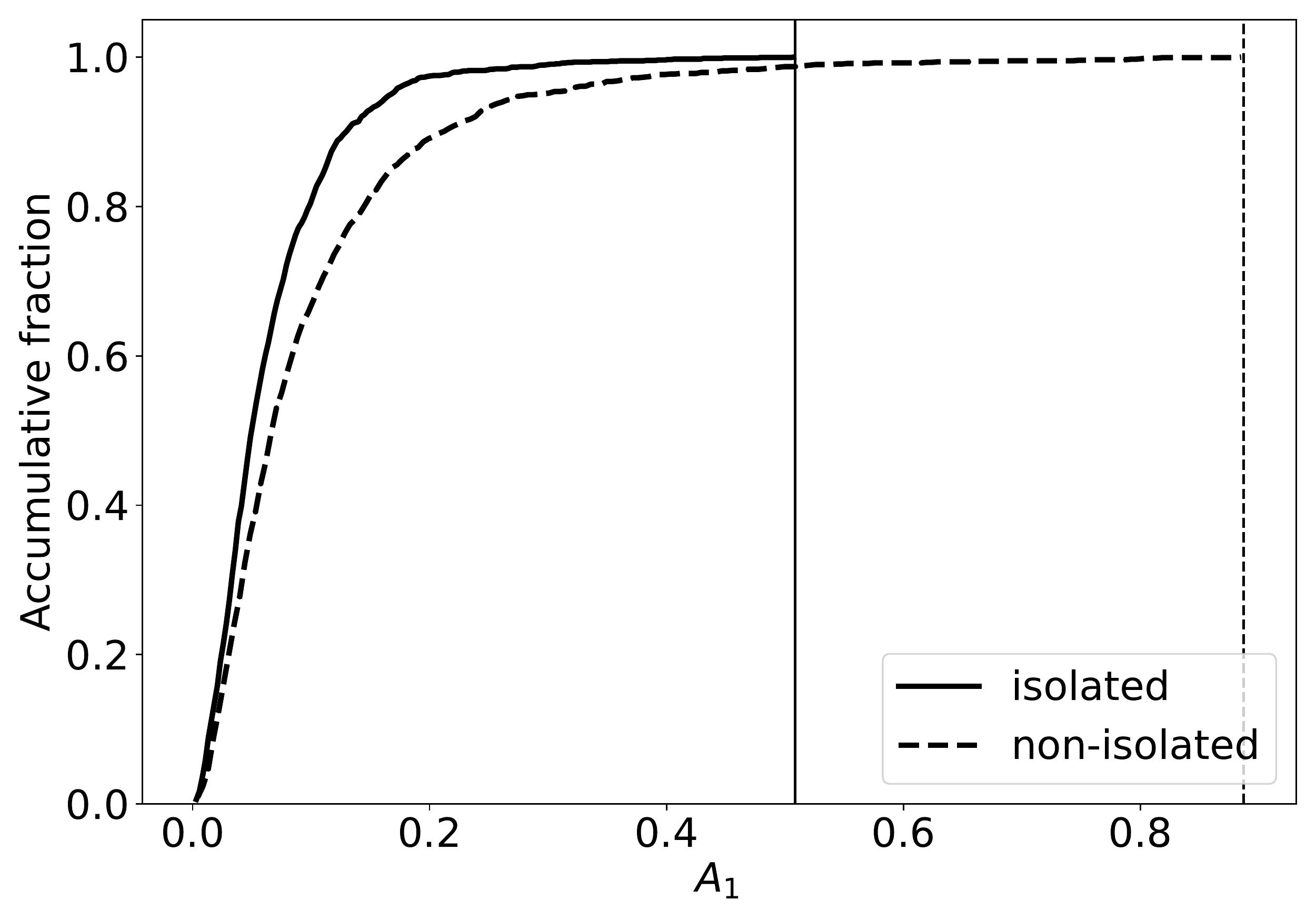}
  \caption{The accumulative fraction of satellite galaxies with different values of m=1 Fourier mode, $A_1$. The two vertical lines show the maximal values of $A_1$ for the isolated and non-isolated samples.} 
    \label{A1}
\end{figure}

In Table \ref{tab:table_1}, the values of $\ar$ for subsamples of isolated and non-isolated galaxies in different mass bins and radius bins are assumed to follow a Gaussian distribution. From the values of $\mu$ and $\sigma$, it is possible to determine whether each pair of subsamples, which share the same mass and radius bins, are from the same distribution. However, it is undetermined if the values of $\ar$ in these subsamples align with a Gaussian distribution, especially for those subsamples with large values of skewness. The Anderson-Darling two-sample tests \citep{Anderson-Darling1952} provide a way to examine whether two samples of data are from the same distribution without a pre-assumption of a specific distribution. Moreover, the Anderson-Darling two-sample test is tail-weighted and thus is more sensitive than the commonly used Kolmogorov-Smirnov test \citep{KolmogorovSmirnov1933}. The null hypothesis that two samples are from the same distribution is rejected when the p-value is larger than 0.05 in an Anderson-Darling 2-sample test.

We perform Anderson-Darling 2-sample tests to investigate whether the distributions of semi-axial ratios $\ar$ are different between the isolated and non-isolated galaxies within the three radial bins for the same mass range. In Table \ref{tab:table_1}, the last column shows the p-values of Anderson-Darling two-sample tests \footnote{We use the {\it twosamples} package from the CRAN R project, which is a statistical computing project. The URL for the {\it twosamples}  package is htpps://github.com/cdowd/twosamples .} for the distributions of $\ar$ of nine pairs of the isolated and non-isolated galaxies binned by their masses and radii. For five of the pairs of galaxy subsamples, the p-values are below 0.05, indicating a deviation from the same distribution of $\ar$. However, p-values are greater than 0.05 for the rest four pairs, which suggests the isolated and non-isolated subsamples with the same mass bin and with the same radius bin may be from the same $\ar$ distribution. Our findings agree with the previous analysis on values of $\mu$ and $\sigma$. For instance, in the first row of Table \ref{tab:table_1}, the values of $\sigma$ from a Gaussian fitting are quite different for the isolated and non-isolated subsamples. This indicates that the two subsamples are not from the same Gaussian distribution. The p-value obtained from the Anderson-Darling two-sample test is 0.0005, indicating a null rejected result. Thus the two subsamples are not from the same distribution without a pre-assumption of a Gaussian form.

Further, let's consider the whole sample (including the above three subsamples) of isolated satellites. We find only $104$ significantly lopsided galaxies whose semi-axial-scale-length deviation to the near side semi-axis, $\Delta a \equiv \frac{|a_--a_+|}{a_+} >0.1$, within $2\re$. That means among the isolated sample of satellites, only $5.8\%$ are significantly lopsided.

Now let's consider the data of $\ar$ ungrouped by masses of galaxes for the isolated and non-isolated samples. Unbinned methods may potentially reveal nuanced effects. Here we examine whether the two unbinned samples of data follow the same distribution in 2D space of $\ar$ and galactic stellar masses. We used the nonparameter two-sample test \citep[][]{Rizzo2019} with bootstrap probabilities to perform the test \citep{Feigelson_Babu2012} by using the {\it cramer} package in the CRAN R project. \footnote{The URL for this package is https://CRAN.R-project.org/package=cramer} The p-values are 0.0030, 0.0050 and 0.0020 within the radial ranges of $0$-$2\re$, $2$-$5\re$ and $5$-$10\re$. Thus the $\ar$ and galactic stellar masses of the isolated and non-isolated sample of galaxies are not from the same distribution.

In the above analysis, $\ar$ is used to define the lopsidedness, which is different from that adopted in the previous analysis based on disc galaxies obtained from the TNG100-1 simulation by \citet{Lokas2022}. It is helpful to make a comparison between our results with theirs. To directly compare the lopsidedness with that in the discs sample \citep{Lokas2022} in TNG-100 simulation, we also calculated the Fourier m=1 mode values, $A_1$, for the isolated satellites on the $x'-y'$ planes in their eigenframes of inertia tensor. The $A_1$ values of satellites are calculated in a radial bin within $1.0-2.0\re$, which is the same as the range of radius used in \citet{Lokas2022}.
We show the accumulative fraction of satellites with growing $A_1$ values in Fig. \ref{A1} and also in Table \ref{tab:table_2}. The fraction of strongly lopsided satellites with $A_1 >0.2$ is $2.5\%$, which is almost an order of magnitude larger than that ($0.37\%$) in \citet{Lokas2022}. One of the possible reasons might be that the sample selections are different. Our sample includes all isolated satellites in a broader stellar mass range of $10^8$-$10^{11}\msun/h$, whereas the sample of \citet{Lokas2022} contains only disc galaxies with stellar masses beyond $10^{10}\msun$. To confirm this, we have further calculated the $A_1$ values for galaxies with the same mass range of \citet{Lokas2022}. There are 876 such galaxies in the Illustris TNG-50 simulation, among which there are $3.5\%$ galaxies moderately lopsided ($A_1 >0.1$), and only one galaxy ($0.1\%$) is strongly lopsided with $A_1 >0.2$. Thus in the same range of masses, the fraction of strongly lopsided galaxies obtained from our analysis agrees with the result from \citet{Lokas2022}, that the simulated satellite galaxies do not exhibit significantly asymmetric.

In addition, in order to examine whether the decreased lopsidedness in our samples, relative to observational samples, is due to our particular choice of satellite galaxies, we calculate the $A_1$ values for a sample of 4001 central galaxies with stellar masses $>10^8\msun$ in the TNG 50-1 simulation. The mean values of $A_{1}$ are $0.059$, $0.048$, $0.032$ and $0.032$ for galaxies within the stellar mass ranges of $10^{8}$-$10^{9}M_{\odot}/h$, $10^{9}$-$10^{10}M_{\odot}/h$, $10^{10}$-$10^{11}M_{\odot}/h$, and greater than $10^{11}M_{\odot}/h$, respectively. Compared to the satellite galaxies, the simulated central galaxies are even less lopsided, in general. 

 Although the fraction of strongly lopsided satellites in a broader range of masses obtained by us for $A_1>0.2$ is one order of magnitude larger than that of \citet{Lokas2022}, it is still much smaller than the observed one \citep{vanEymeren+2011,Yu+2022}. It is noteworthy that the stellar mass ranges of our work and of \citet{Yu+2022} are very similar, while the stellar masses are $\in[10^8,~10^{11.5}]~\msun$ with a median of $10^{9.7}\msun$ in the latter case. Considering the lopsidedness decreases as stellar mass increases in the above studies, it seems unlikely the small fraction of galaxies with stellar masses of $10^{8}\msun$-$10^{8} \msun/h$ or of $10^{11}\msun /h$-$10^{11.5}\msun$ contribute to the majority of the proportion of asymmetric galaxies, making it statistically consistent with observations.

\begin{figure}
  \includegraphics[width=\columnwidth]{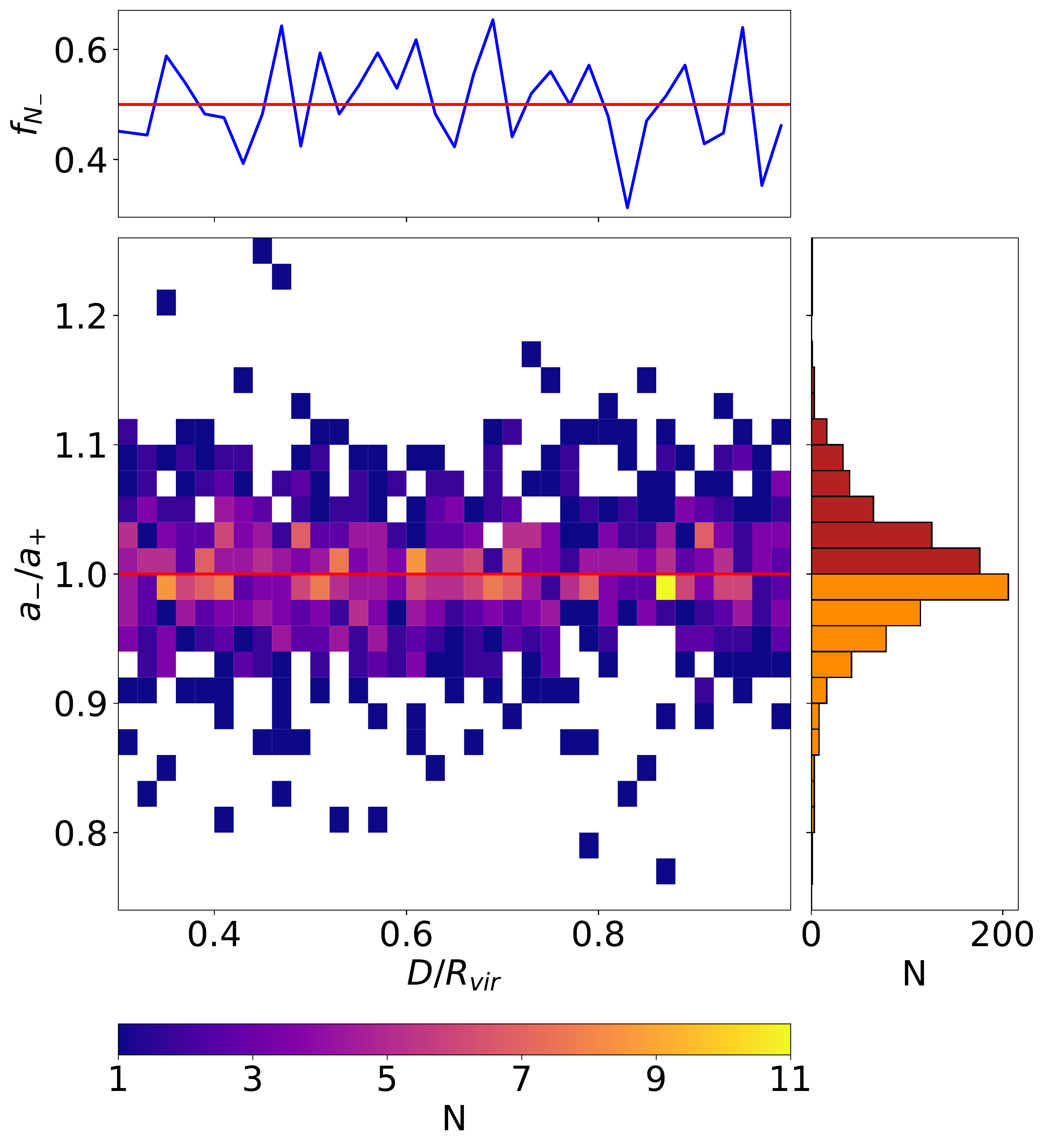}
  \caption{The lower left panel: The correlation between semi-axial ratios, $\ar$ at $2\re$, for the 1780 present-day isolated satellites in $3\sigma$ of Gaussian fitting and the distances, $D$, of satellites to their host halo centres. The 2-dimensional histograms of data are binned in $0.02 \rvir$ intervals in distance and $0.01$ intervals in $\ar$, and are represented using the colour bars. The ratio $\ar=1.0$ is shown with a horizontal line, as well. The fractions of satellites with $\ar<1.0$ at all distances are presented in the upper left panel and there is a histogram of satellites with different values of $\ar$ in the right panel.
  }\label{fig:fig_3}
\end{figure}

\begin{figure}
  \includegraphics[width=\columnwidth]{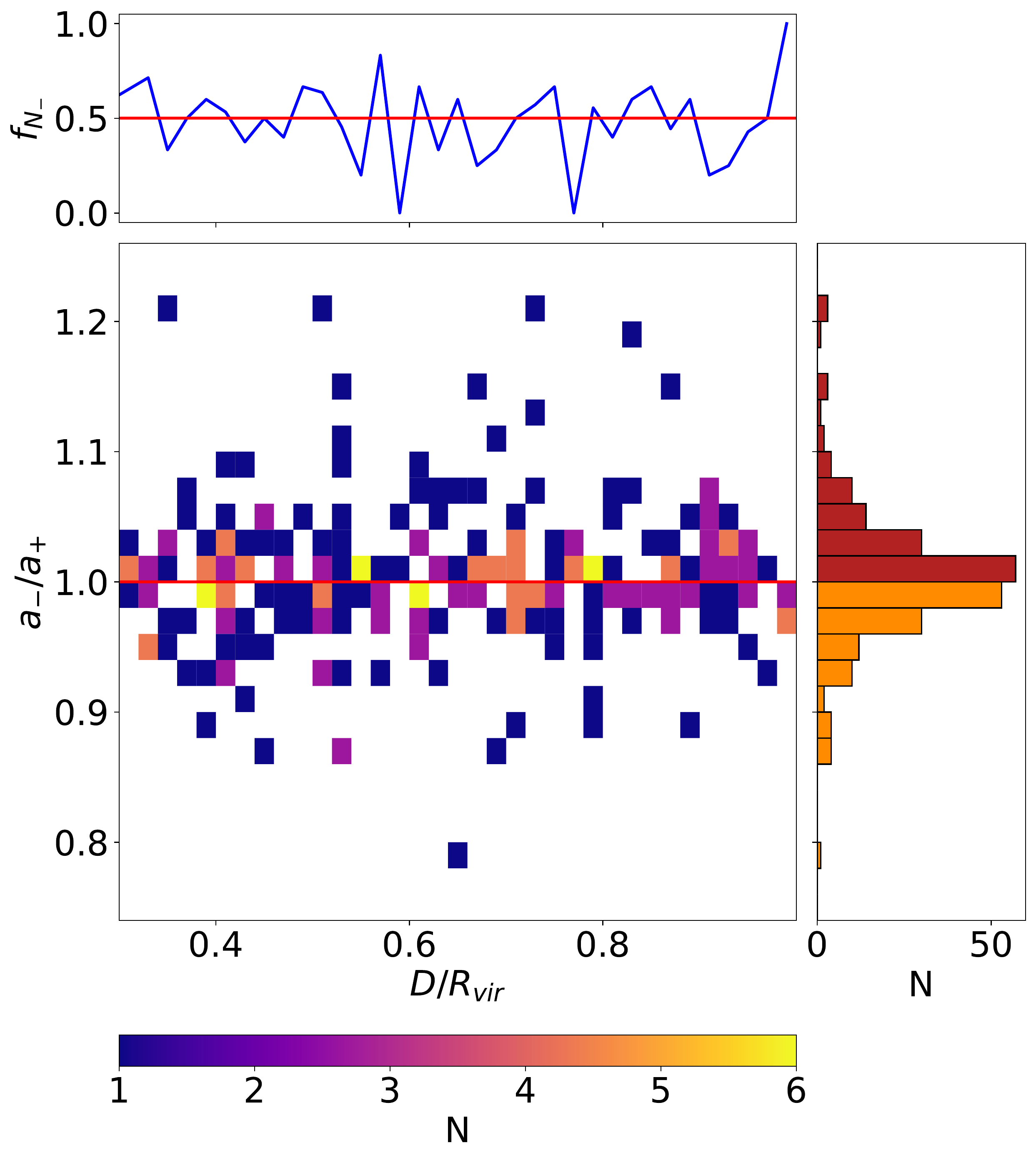}
  \caption{The lower left panel: The correlation between semi-axial ratios, $\ar$ at $10\re$, for the present-day isolated satellites in $3\sigma$ of Gaussian fitting and the distances, $D$, of satellites to their host halo centres. The binning method is the same as that in Fig. \ref{fig:fig_3}. The fractions of satellites with $\ar<1.0$ at all distances are presented in the upper left panel. A histogram of satellites with different values of $\ar$ is given in the right panel.
  }
  \label{fig:fig_4}
\end{figure}

Next, we turn to consider the sample of non-isolated satellites. In this case, the values of $N_-$ are larger than those of $N_+$. Moreover, $\Delta a$ is further enhanced. In addition, we find $180$ significantly lopsided satellites with $\Delta a>0.1$ in the radial range of $0$-$2\re$ out of $1409$ non-isolated satellites. The fraction of significantly lopsided non-isolated satellites is about $12.8\%$ under the criteria of $\Delta a>0.1$, nearly a factor of 2 larger than the fraction in the isolated sample. If $A_1$ is used instead of $\Delta a$, $11.1\%$ of the non-isolated satellites are strongly lopsided with $A_1>0.2$. The larger fraction measured by $A_1$ can be attributed to the usage of a locally radial range of $[1.0,~2.0\re]$ in calculations. It is the asymmetric distribution of stars within the range of $ [1.0,~2.0\re]$ that makes $A_1$ larger. All stellar particles within $2\re$ are included in the calculation of $\Delta a$, so the global asymmetry is weakened if the central regions are symmetric.

In a word, the non-isolated satellites are made more lopsided by the perturbations from a close encounter or a minor merger (see \S \ref{origins}). The fraction of strongly lopsided satellites in our non-isolated sample is much higher than that obtained by \citet{Lokas2022}, but is also lower than that in observations \citet[e.g.,][]{vanEymeren+2011,Yu+2022}, as in the case of isolated satellites.

\begin{figure}
  \includegraphics[width=\columnwidth]{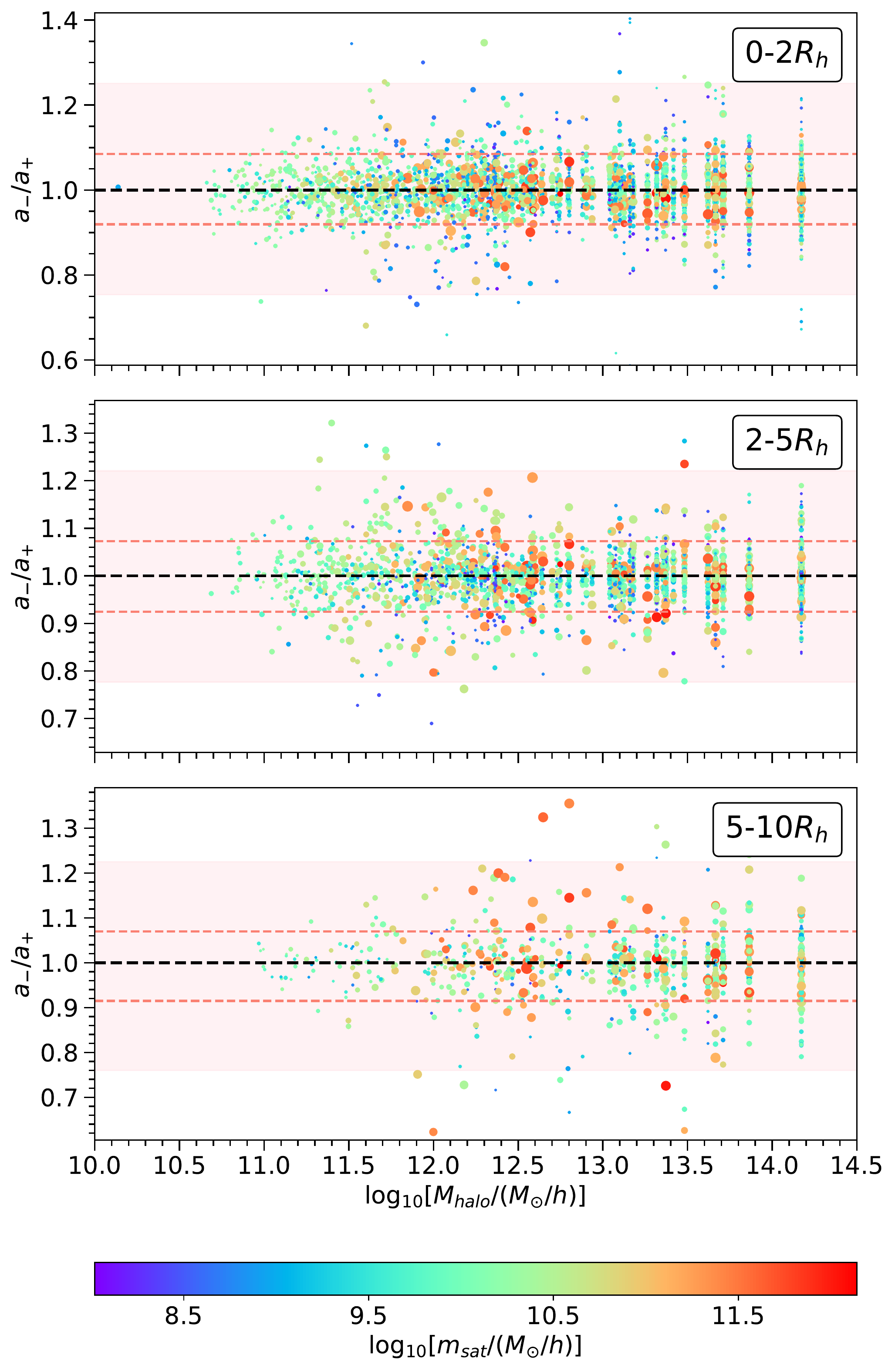}
  \caption{The relation between the semi-axial ratios in the inner (upper panel), intermediate (middle panel) and outer (lower panel) regions of the overall sample of 3189 satellites and their host halo masses. The satellites are presented in colour-filled circles. Different colours and sizes of the circles show the masses and the sizes of the satellites. The axial ratios $\ar=1.0$ are plotted as black dashed lines. The red dashed lines and pink shadows show the areas of $1\sigma$ and $3\sigma$ of $\ar$ in the Gaussian fitting. 
  }
  \label{mass_ar}
\end{figure}

\subsection{Shape correlations}

As aforementioned, the three subsamples of isolated galaxies display weak lopsidedness in their shapes. The long semi-axes of satellites do not preferably face the direction of halo centres. This confirms that the lopsided shapes of satellites are not mainly caused by the tidal fields from the halo centres. To examine the tidal field effects from the halo centres more carefully, we further check the inner and outer semi-axial ratios, $\ar$, in different environments of the haloes, and show the results in Figs. \ref{fig:fig_3}-\ref{fig:fig_4}.

The isolated sample does not contain galaxies near their host central galaxies, say, no isolated galaxies located within $0.3\rvir$ of the halo. The histograms of $\ar$ for the inner (Fig. \ref{fig:fig_3}) and outer (Fig. \ref{fig:fig_4}) regions of satellites show that the asymmetry of the galaxies is weak. The values of $\ar$ for most of the satellites are between $0.9-1.1$, which agrees with the conclusion in Table \ref{tab:table_1}. In the dense regions of satellites, i.e., within $0$-$2\re$ of the satellites, the fraction of satellites with $\ar<1.0$ (hereafter denoted as $f_{N_-}$), oscillates around $0.5$ for the isolated sample of satellites. 
The number of isolated satellite galaxies should be reduced when studying the outer regions of the satellites since we always require more than a thousand stellar particles in the radial ranges of a satellite studied. As shown in Fig. \ref{fig:fig_4}, the conclusion that asymmetry is absent in the isolated sample is still valid in the outer regions ($5$-$10\re$). In summary, the tidal fields from halo centres do not contribute to the lopsidedness of the satellites.

The Jacobi radius of a satellite on a circular orbit is approximately \citep{BT2008}
\beq\label{rtidal}
\tide=D\left(\frac{M_{\rm sat}}{\zeta M_{\rm halo}}\right)^{1/3}, ~~~\zeta=3 \eeq
and the tidal radii on other kinds of orbits have the same form but with different values of $\zeta$.
The lopsidedness is not directly related to the tidal force from halo centres for most of the satellites outside $0.2\rvir$. An implication is that there are little correlations between the physical quantities appearing in Eq. \ref{rtidal} ($M_{\rm sat}$, $M_{\rm halo}$ and $D$) and the lopsidedness. Note that $M_{\rm sat}$ includes both baryonic and subhalo dark matter masses of a satellite galaxy. We use the results shown in Fig. \ref{mass_ar} to confirm that in all truncation radii of satellites, the semi-axial ratios $\ar$ do not correlate with the halo masses nor the overall satellite masses.

\begin{figure}
  \includegraphics[width=80mm]{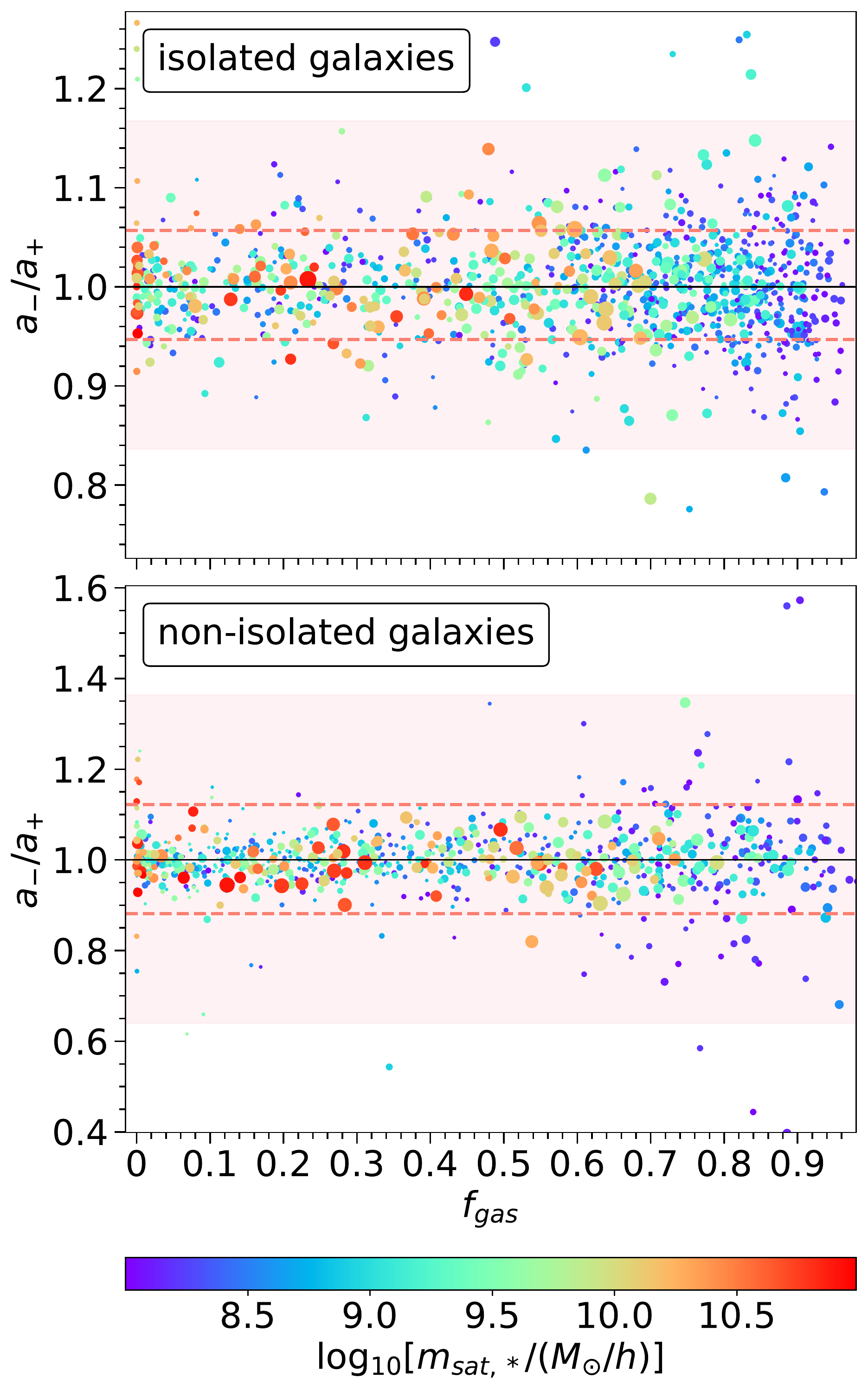}
  \caption{A relation between $\ar$ and the fraction of gaseous component in the isolated sample (upper panel) and non-isolated sample (lower sample). The satellites are truncated at $2\re$ in the calculations of $\ar$. The red dashed lines and pink shadows show the areas of $1\sigma$ and $3\sigma$ of $\ar$ in the Gaussian fitting.}
    \label{gasfrac}
\end{figure}

Observations have revealed that the HI distribution in galaxies is typically asymmetric \citep[e.g.,][]{Watts+2020a}. In simulations, the HI distribution of satellite galaxies in TNG100-1 is generally asymmetric \citep{Watts+2020b}. Recently, \citet{Lokas2022} pointed out the lack of a direct relationship between the global asymmetries of the stellar and the gaseous components in the TNG100 simulation. Below, we study the relation between the gas fractions and the lopsidedness of satellite galaxies. The values of $\ar$ for satellites truncated at $2\re$ versus the fraction of gas, $f_{\rm gas}$ in a satellite, are plotted in Fig. \ref{gasfrac} for isolated (upper panel) and non-isolated (lower panel) satellites. The fraction of gas is defined by $f_{\rm gas}=\frac{m_{\rm sat,gas}}{m_{\rm sat,b}}$, where $m_{\rm sat,gas}$ and ${m_{\rm sat,b}}$ are the masses of gas and baryons of a satellite galaxy. In both samples, there is a clear signal that more massive satellites tend to be less gas-rich, and the dispersion of $\ar$ is smaller around 1.0. More massive satellites are more symmetric with smaller values of $\Delta a$. As satellite mass decreases, $\Delta a$ becomes larger. Satellites with a larger fraction of gaseous components appear to be more lopsided.

\section{Alignments}\label{alignment}
\begin{figure*}
  \includegraphics[width=170mm]{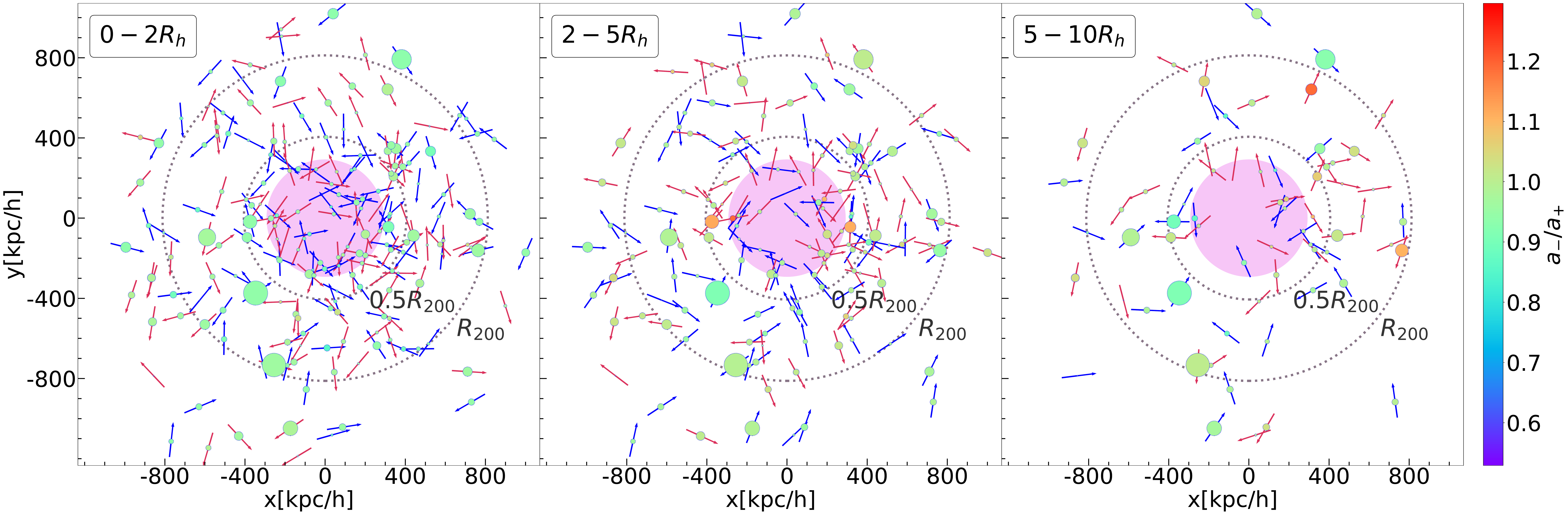}
  \caption{The spatial distribution and orientation of the major axes of low mass satellites in the most massive halo, Halo 0, viewed along the $z$-direction obtained directly in TNG50-1 simulation. The whole halo is recentred and the original point is the halo centre. The central pink region shows the radius of the central galaxy within $10\re$. The two dashed circles indicate the radii of $0.5\rvir$ and $\rvir$ of the halo. The satellites with different semi-axial ratios are plotted in different colour circles. The radii of the circles are $10\re$ of the satellites. The solid lines going through the satellite circles indicate their major axes, blue for the inwards and red for the backward directions in $3-$dimensional space to the halo centres for the long semi-axes, together with the arrows.} 
    \label{halo0RA}
\end{figure*}

\begin{figure}
  \includegraphics[width=90mm]{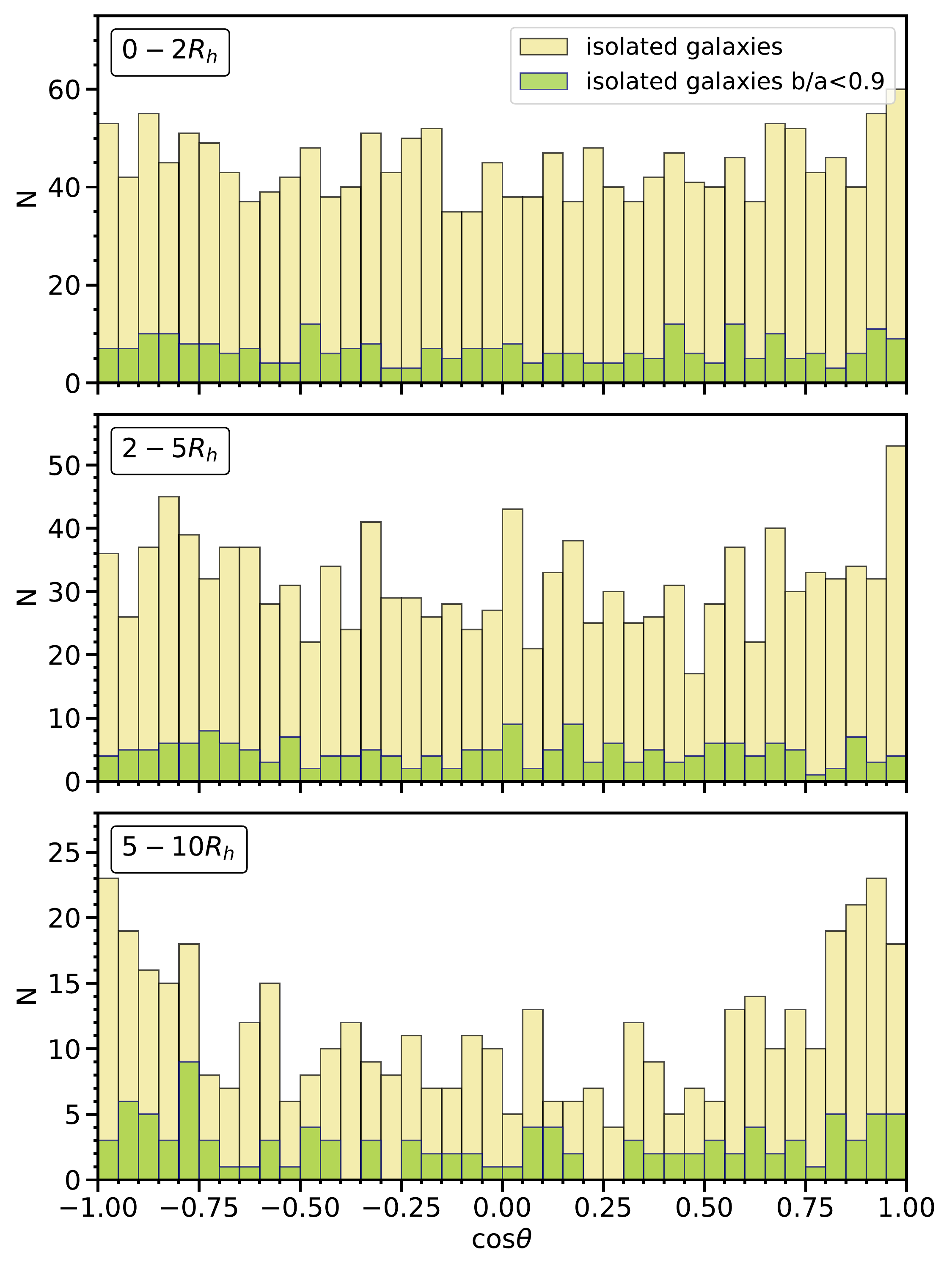}
  \caption{The histograms of the cosines of aligned angles between the long semi-major-axes of satellites and their halo centres in the inner regions within $0$-$2\re$ (upper panel), intermediate regions within $2$-$5\re$ (middle panel) and outer regions within $5$-$10\re$ (lower panel). The $\cos\theta$ are binned in $0.05$ for the overall sample of isolated satellites (the flavogreen histograms) and the triaxial isolated satellites with intermediate-to-long axial ratios $b:a<0.9$ (the green histograms).} 
    \label{theta}
\end{figure}

The existence of RA is still controversial. The RA between the major axis of a satellite and the radial direction of its central halo was found in observations in the 1970s \citep{Hawley_Peebles1975, Thompson1976}. More recently, some observations further confirmed the existence of the RA in large samples of galaxies \citep{Pereira_Kuhn2005,Faltenbacher+2007a,Wang+2008,Wang+2019}, but other observations do not find evidence of RA at different redshifts \citep{Hung_Ebeling2012,Schneider+2013,Chisari+2014,Sifon+2015}. On the other hand, RA was discovered in the cosmological simulations by \citet{Faltenbacher+2008,Pereira+2008,Knebe+2020}. Especially, it is more pronounced on the small scale of $1.5\rvir$ \citep{Faltenbacher+2008}. The mean value of the alignment angle, $\phi$, i.e., the angle between the major axis of a satellite and the radial direction of the halo centre, is adopted to determine the existence of RA in \citet{Faltenbacher+2008}. The long and short semi-major-axes are not distinguished in the above studies, thus the values of $\phi$ are taken to lie in $[0^\circ,~90^\circ]$. For a sample of satellites radially aligned with their halo centres, the mean value of $\phi$ is less than $45^\circ$. In addition, besides the radial alignment of satellites, \citet{Wang+2019} observed a tangential alignment in the satellites around bound primary galaxy pairs in the SDSS DR13. There is a clear trend that the major axes of satellites located in the area between two primary galaxies are perpendicular to the radial direction of their host halo. The tangential alignment is beyond the scope of this work, and we shall not further discuss this issue.

The RA of isolated satellites with a mass of $10^8$-$10^{9}\msun /h$ in the most massive halo (Halo 0) in TNG50-1 is qualitatively shown in Fig. \ref{halo0RA}. These satellites are the ones most substantially perturbed by the tides from their host halo. The three panels present the RA in all chosen truncation radii of satellites. The orientations of long semi-major axes of satellites are plotted with blue (inwards) and red (backward) arrows on the circles of satellites. We observe no significant RA in the radial bins of $0$-$2\re$, $2$-$5\re$ and $5$-$10\re$ at the first glance of the figure. 
To examine whether RA exists or not, below we perform a more careful analysis.

\subsection{RA and LRA}
As aforementioned, it remains unclear whether LRA can be produced in the simulated galaxies within the $\Lambda$CDM framework. Here we provide the first systematical analysis of the existence of LRA in the sample of satellite galaxies generated from the TNG50-1 simulation.

We need to consider the RA before analysing LRA. For this purpose, here we define a 3D alignment angle, $\theta$, as follows,
\beq\label{aa} \cos\theta = {\rm \bold x'_{sat}} \cdot {\rm \bold D},\eeq
where ${\rm \bold x'_{sat}}$ is the eigenvector of the long semi-major axis of a satellite, and ${\rm \bold D}$ is the radial direction to the halo centre. The values of $\cos\theta$ are calculated for the sample of isolated satellites. The definition of the 3D alignment angle in the existing work of alignments in galaxies is in the range of $[0^\circ, ~90^\circ]$ \citep[e.g.,][]{Libeskind+2009,Velliscig+2015,Tempel+2015,Zhang+2023} under the assumption of triaxially symmetric satellite galaxies. Given that the long and short semi-major axes are distinguishable in the lopsided galaxies in this work, $\theta$ takes all the possible values within $[0^\circ,~180^\circ]$. As a result, $\cos\theta \in [-1,1]$. For a satellite with $\cos\theta \in [0,1]$, the long semi-major axis is radially aligned with the direction of the halo centre, and for $\cos\theta \in [-1,0]$, the short semi-major-axis is radially aligned. 

The values of $\theta$ can be used to determine the existence of RA and the LRA. It is convenient to use ${\langle\cos\theta_1\rangle}$ and  ${\langle\cos\theta_2\rangle}$ to denote the mean values of $\cos\theta$ in the ranges of $ [0,1]$ and $[-1,0]$, respectively. An RA emerges when ${\langle\cos\theta_1\rangle}>0.5$ or ${\langle\cos\theta_2\rangle} <-0.5$. Moreover, LRA refers to the special case in which the long semi-major axis of a lopsided satellite radially aligns with the halo centre, corresponding to $\cos\theta>0$. The existence of LRA can be determined by the $\cos\theta$-dependence of the number of satellites.

\subsection{On the existence of RA}

In Fig. \ref{theta}, the values of $\cos\theta$ are binned in $0.05$ in the inner, intermediate and outer regions for the overall sample of isolated galaxies with flavogreen histograms. The alignment signal shows a strong dependence on the radial ranges. For the radial ranges of $0$-$2\re$ and $2$-$5\re$, the number of galaxies in each bin exhibits a relatively random $\cos\theta$-dependence, without discernible anisotropy. To investigate whether the distributions of $\cos\theta$ for the isolated sample in each radius bin follow the uniform distribution, we perform the Anderson-Darling 2-sample tests with random numbers following a uniform distribution within $[-1, ~1]$. The resulting p-values are 0.224 and 0.089 in radial bins of $0-2\re$ and $2$-$5\re$, suggesting that the distributions of $\cos\theta$ in these two radial ranges may follow uniform distributions. In the radius bin of $5$-$10\re$, the number distribution of galaxies appears as a U-shape distribution, and the AD test with a uniform distribution returns a p-value of 0.0005. Apparently, the distribution of 3D alignment angles is not uniform. Such a U-shape distribution is compelling evidence for the existence of radial alignment. 

We calculate the values of ${\langle\cos\theta_1\rangle}$ and ${\langle\cos\theta_2\rangle}$ at all truncation radii for the isolated galaxies binned in radius. 
We find that ${\langle\cos\theta_1\rangle=0.52}$ and ${\langle\cos\theta_2\rangle=-0.51}$ for $0$-$2\re$, ${\langle\cos\theta_1\rangle=0.52}$ and ${\langle\cos\theta_2\rangle=-0.53}$ for $2$-$5\re$ and ${\langle\cos\theta_1\rangle=0.61}$ and ${\langle\cos\theta_2\rangle=-0.58}$ for $5$-$10\re$. The mean values of the misalignment angle agree with the Anderson-Darling tests. 
A statistically significant signal of the RA emerges in the outer region of galaxies, which is in agreement with previous results obtained in cosmological simulations \citep{Kuhlen+2007,Pereira+2008,Knebe+2020}. Contrarily, the presence of RA is not clearly observed in the inner and intermediate regions. Although the mean values of ${\langle\cos\theta_1\rangle} >0.5$ and ${\langle\cos\theta_2\rangle}<-0.5$, the differences between these values and the mean value of a uniform distribution so small that it is impossible to confirm the existence of RA in the radial bins of $0$-$2\re$ and $2$-$5\re$. 

In the samples of  \citet{Pereira+2008,Knebe+2008,Knebe+2020}, triaxial galaxies are selected. Our sample includes both triaxially and axially symmetric satellites. For an oblate elliptical satellite or a disc satellite, the major and intermediate axes are indistinguishable. For such an oblate/disc galaxy, the value of $\cos\theta\approx 0$ might arise due to the intermediate axis being identified as the major axis, and therefore a well-aligned galaxy is artificially classified as misaligned. To avoid this problem, we remove the nearly axially symmetric satellites (with an axial ratio $b:a \in [0.9,~1.0]$) from our sample, following the strategy used by \citet{Pereira+2008,Knebe+2008} and \citet{Knebe+2010}. After doing so, the rest subsample contains only triaxial systems. \footnote{The irregular satellites are also considered to be triaxially symmetric.} The $\cos\theta$ values of the triaxial satellites are computed and presented in Fig. \ref{theta} with green histograms. ${\langle\cos\theta_1\rangle}=0.53$ and ${\langle\cos\theta_2\rangle}=-0.53$ within $0$-$2\re$, ${\langle\cos\theta_1\rangle}=0.46$ and ${\langle\cos\theta_2\rangle}=-0.54$ within $2$-$5\re$ and ${\langle\cos\theta_1\rangle}=0.58$ and ${\langle\cos\theta_2\rangle}=-0.61$ within $5$-$10\re$. The mean values of the misalignment angles suggest a clear signal of RA only appears in the outer regions of the new subsample of triaxial galaxies.
Therefore the absence of RA in the inner and intermediate regions is not caused by the axial symmetry of satellite galaxies. However, our findings in the inner and intermediate regions do not conflict with the earlier studies \citep[such as][]{Kuhlen+2007,Pereira+2008,Faltenbacher+2008,Knebe+2020}. In some of those studies, despite satellite galaxies, the RA between subhalos and their host halos are explored \citep[e.g.,][]{Kuhlen+2007,Pereira+2008}. The dark matter subhalos can extend to a radius much further than $5\re$ of the stellar component. Furthermore, in the aforementioned research, even when satellite galaxies are recognized as the stellar components of subhalos, the radial alignment between these satellites and their host halos is examined by considering all stellar particles \citep[e.g.,][]{Knebe+2020}. In fact, the faint signal of RA in the inner and intermediate regions in this work aligns with the results found by \citet{Faltenbacher+2008}. The trend of decreasing ${\langle\cos\theta_1\rangle}$ or increasing ${\langle\cos\theta_2\rangle}$ as the truncation radius decreases is consistent with the earlier analysis \citep[e.g.,][]{Kuhlen+2007}.

In conclusion, our research indicates that RA only appears in the outermost regions of isolated galaxies in the TNG 50-1 simulation, a region where tidal interactions from the centres of host halos are prominent. This could potentially suggest a tidal origin for the occurrence of RA, which is consistent with the previous analysis based on other cosmological simulations.

\subsection{On the existence of LRA}\label{LRACDM}
For a sample of satellites, LRA exists if there is a significantly larger number of satellites with $\cos\theta>0$ than that with $\cos\theta<0$. Note that the existence of LRA is determined by ${\langle\cos\theta\rangle}>0$ rather than ${\langle\cos\theta\rangle}<0.5$ for the whole sample, since a dominant fraction of the satellites is axial-symmetric (as seen from Fig. \ref{theta}). Fig. \ref{theta} shows almost mirror-symmetric distributions, centred at $\cos\theta=0$, of the numbers of isolated satellites (flavogreen histograms) in the radial ranges of $0$-$2\re$ and $2$-$5\re$. The average values of the $\cos\theta$ in the inner and intermediate regions are $0.003$ and $-0.005$, respectively. Moreover, the skewness of $\cos\theta$ are $0.002$ and $0.042$ in these radial bins. The mean value and skewness indicate that the long semi-major axes of the satellites in these radii do not prefer to point towards the direction facing their halo centres, i.e., the near side. Thus the LRA is not found in the isolated satellites in the inner and intermediate regions.

In the outer regions of $5$-$10\re$, the distribution of the numbers of satellites is different. The histogram exhibits a random $\theta$-dependence, and ${\langle\cos\theta\rangle}=0.002$ with a skewness of $0.008$. Thus the LRA is not observed in the outer regions, either. The absence of the LRA agrees with our previous findings in \S \ref{axialratio}. In summary, the longer side of the major axis of a lopsided isolated satellite galaxy does not preferentially point towards the host halo centre. As aforenoted, the tidal fields induced by the host halos may be the origin for the existence of RA in the TNG 50-1 satellite galaxies, since the RA emerges in the outskirts of these galaxies where tidal fields are substantial. 

The lack of LRA points out that the far-distance tidal fields triggered by the host halos do not lead to detectable systematic signals of lopsidedness for the satellites. However, the lopsidedness of galaxies may still be caused by short-distance tidal forces generated by nearby massive objects such as a massive galaxy. For a satellite galaxy moving along an extremely radial orbit, the mass distribution of the satellite might be tidally reshaped at its pericentre. For instance, if the satellite is close enough to the massive galaxy, the outskirts of the galaxy fill the tip and the broad ends of the Roche lobes. Thus the galaxy is lopsided in morphology. When the galaxy is moving radially outwards the massive galaxy, the lopsided shape can exist for a longer time scale than the orbital time. Moreover, since the majority of the satellites are disc or oblate galaxies, they are rotationally-supported systems. The rotation of the satellites might change the orientations of the long semi-major axes \citep[][]{Mapelli+2008} and thus may erase the signals of LRA.

\section{Discussions}\label{discussions}

\subsection{Extremely lopsided satellites and absence of LRA}\label{extremecases}

\begin{table*}
  \centering
  \caption{Extremely lopsided satellites with semi-axial ratios within the radial range of $0$-$2\re$, $(\ar)_{2\re}$, out of 3$\sigma$ of the Gaussian fitting. The information on isolated and non-isolated satellites is shown in the upper and the lower sub-tables, respectively. The IDs of the subhaloes within which the satellites are embedded are listed in the first column. The IDs of the host haloes for the satellites are listed in the second column. The masses of the stellar, gaseous and dark matter of the satellites are provided in the third to the fifth columns. The half-mass radii and tidal radii of satellites are itemised in the sixth and eighth columns, together with the distances between the satellites and their host halo centres in the seventh column. The semi-axial ratios of the major axes in the eigenframe of the satellites calculated within $0$-$2\re$, and the values of the $m=1$ Fourier mode, $A_1$, computed within a local radial range of $1$-$2\re$, are presented in the ninth and tenth columns, respectively. }
  \label{tab:table_3}
  \begin{threeparttable}
  \begin{tabular}{lcccccccccccccr} 
    \hline
    subID  & haloID & $m_{\rm sat,*}$ & $m_{\rm sat,gas}$ & $m_{\rm sat,dm}$ & $R_{h}$ & $D$  & $\tide$ & $(a_-/a_+)_{2\re}$ & $(A_1)_{2\re}$\\
    & & $(10^{8}M_{\odot}/h)$ & $(10^{8}M_{\odot}/h)$ & $(10^{8}M_{\odot}/h)$ & $({\rm kpc}/h)$ & $(\rvir )$ & $({\rm kpc}/h)$ &\\
    \hline \multicolumn{10}{c}{Isolated satellites with extreme lopsidedness} \\
    \hline
    96819 & 2 & 3.0 & 0.0 & 137.0 & 3.4 & 0.73 & 18.88 & 1.19 & 0.13\\
    282875 & 15 & 1.5 & 0.0 & 0.7 & 0.5 & 0.56 & 3.55 & 0.80 & 0.36\\
    429476 & 68 & 10.8 & 12.2 & 169.7 & 2.1 & 0.36 & 9.89 & 1.20 & 0.45\\
    460724 & 95 & 72.3 & 168.3 & 570.4 & 4.3 & 0.78 & 36.18 & 0.78 & 0.12\\
    108 & 0 & 31.9 & 0.0 & 0.0 & 0.2 & 0.37 & 5.84 & 1.37 & 0.39\\
    320 & 0 & 5.9 & 0.0 & 4.2 & 0.3 & 0.56 & 5.96 & 0.69 & 0.51\\
    96782 & 2 & 83.7 & 0.0 & 493.5 & 0.9 & 0.46 & 18.84 & 1.24 & 0.43\\
    96804 & 2 & 31.0 & 0.0 & 82.8 & 0.6 & 0.41 & 9.90 & 0.82 & 0.43\\
    96883 & 2 & 5.6 & 0.0 & 0.0 & 0.2 & 0.33 & 2.91 & 0.82 & 0.30\\
    143916 & 4 & 1.7 & 1.6 & 218.5 & 2.8 & 1.34 & 34.20 & 1.23 & 0.10\\
    184943 & 6 & 150.4 & 0.1 & 154.3 & 0.8 & 0.57 & 18.57 & 1.27 & 0.40\\
    185048 & 6 & 1.0 & 0.0 & 26.3 & 1.1 & 1.07 & 15.57 & 0.84 & 0.20\\
    282802 & 15 & 10.5 & 0.0 & 5.3 & 0.4 & 0.54 & 6.51 & 0.82 & 0.32\\
    300908 & 18 & 15.1 & 77.3 & 535.7 & 3.3 & 1.16 & 48.13 & 1.25 & 0.01\\
    433290 & 71 & 4.1 & 6.5 & 214.7 & 1.2 & 0.84 & 25.91 & 0.83 & 0.31\\
    482891 & 120 & 44.1 & 0.0 & 23.1 & 0.5 & 1.26 & 26.13 & 1.21 & 0.28\\
    507786 & 157 & 5.3 & 16.1 & 204.3 & 1.1 & 0.87 & 23.24 & 0.77 & 0.35\\
    529366 & 198 & 18.7 & 37.7 & 289.5 & 2.0 & 0.87 & 30.79 & 1.50 & 0.48\\
    564827 & 296 & 3.0 & 13.8 & 273.8 & 1.1 & 1.57 & 49.17 & 0.80 & 0.29\\
    567005 & 303 & 8.2 & 40.5 & 468.2 & 1.5 & 2.13 & 79.41 & 1.25 & 0.38\\
    581059 & 356 & 3.4 & 50.0 & 444.3 & 1.4 & 2.61 & 97.97 & 0.79 & 0.41\\
    584877 & 372 & 9.6 & 26.0 & 158.8 & 0.9 & 0.48 & 13.85 & 1.25 & 0.19\\
    585518 & 375 & 4.3 & 33.1 & 247.8 & 2.4 & 1.23 & 38.26 & 1.19 & 0.17\\
    \hline \multicolumn{10}{c}{Non-isolated satellites with extreme lopsidedness} \\
    \hline
    365 & 0 & 5.5 & 0.0 & 0.0 & 0.2 & 0.71 & 0.02 & 1.53 & 0.53\\
    1121 & 0 & 1.4 & 0.0 & 0.0 & 0.06 & 0.71 & 0.01 & 1.50 & 0.48\\
    63990 & 1 & 11.0 & 0.0 & 4.8 & 0.3 & 0.03 & 0.42 & 1.32 & 0.39\\
    185045 & 6 & 1.3 & 5.7 & 0.0 & 3.3 & 2.07 & 2.22 & 2.19 & 0.63\\
    275577 & 14 & 1.8 & 13.6 & 0.0 & 1.3 & 0.10 & 1.23 & 1.52 & 0.68\\
    275579 & 14 & 9.2 & 0.0 & 0.0 & 0.2 & 0.16 & 0.02 & 1.40 & 0.37\\
    275622 & 14 & 1.6 & 5.2 & 0.0 & 0.9 & 0.10 & 0.95 & 0.58 & 0.75\\
    300911 & 18 & 41.2 & 3.0 & 0.0 & 0.1 & 0.07 & 0.07 & 0.64 & 0.40\\
    300972 & 18 & 4.4 & 0.0 & 0.0 & 0.03 & 0.07 & 0.01 & 2.32 & 0.76\\
    342487 & 26 & 1.1 & 4.7 & 0.0 & 1.3 & 1.38 & 0.87 & 1.71 & 0.69\\
    371134 & 36 & 1.3 & 11.9 & 0.0 & 1.5 & 0.12 & 1.29 & 1.57 & 0.52\\
    435761 & 73 & 2.2 & 11.3 & 0.0 & 2.0 & 0.18 & 2.08 & 0.22 & 0.80\\
    435762 & 73 & 1.4 & 10.6 & 0.0 & 1.9 & 0.16 & 1.76 & 0.40 & 0.89\\
    474022 & 109 & 1.2 & 7.6 & 0.0 & 1.4 & 0.10 & 1.08 & 2.29 & 0.82\\
    479954 & 116 & 3.4 & 0.0 & 2.4 & 0.4 & 0.05 & 0.42 & 1.71 & 0.66\\
    510286 & 161 & 1.1 & 5.6 & 0.0 & 1.0 & 0.15 & 1.49 & 0.44 & 0.81\\    
    535781 & 212 & 1.4 & 1.9 & 0.0 & 0.6 & 0.04 & 0.28 & 0.39 & 0.79\\
    603005 & 463 & 8.4 & 4.4 & 15.0 & 1.1 & 0.09 & 1.35 & 0.54 & 0.58\\
    63944 & 1 & 21.1 & 0.0 & 20.1 & 0.5 & 0.11 & 1.83 & 1.31 & 0.37\\
    253885 & 12 & 32.7 & 0.1 & 0.0 & 0.1 & 0.05 & 0.66 & 1.24 & 0.28\\
    452981 & 88 & 38.8 & 114.4 & 159.4 & 3.2 & 0.34 & 11.57 & 1.45 & 0.41\\
    494013 & 135 & 27.0 & 2.7 & 0.2 & 0.2 & 0.04 & 0.63 & 0.66 & 0.24\\
    122 & 0 & 28.7 & 0.0 & 0.0 & 0.2 & 0.14 & 2.08 & 1.26 & 0.36\\
    125 & 0 & 26.5 & 0.0 & 0.0 & 0.2 & 0.14 & 2.13 & 0.67 & 0.50\\
    161 & 0 & 18.9 & 0.0 & 0.0 & 0.2 & 0.25 & 3.33 & 0.72 & 0.45\\
    96783 & 2 & 125.7 & 0.2 & 70.2 & 0.7 & 0.27 & 7.89 & 1.22 & 0.29\\
    117308 & 3 & 33.6 & 0.0 & 0.0 & 0.2 & 0.23 & 3.16 & 1.23 & 0.38\\
    275567 & 14 & 14.9 & 0.0 & 0.0 & 0.2 & 0.18 & 2.13 & 1.39 & 0.38\\
    \hline
  \end{tabular}
  \end{threeparttable}
\end{table*}

The extremely lopsided galaxies with an $\ar$ beyond the $3\sigma$ of the Gaussian fitting are listed in Table \ref{tab:table_3} for the isolated sample (the upper sub-table) and the non-isolated sample (the lower sub-table).

Eighteen out of 23 extremely lopsided non-isolated satellites are strongly influenced by the tidal fields from the halo centres or nearby massive galaxies. As a result, the tidal radius $\tide$ is smaller than $2\re$. There are seven satellites with $\ar<1.0$ and eleven with $\ar>1.0$ among the satellites close to halo centres or nearby massive galaxies. These galaxies are listed in the $1^{st}-18^{th}$ rows of the lower sub-table in Table \ref{tab:table_3}.
There are no such strongly tidally perturbed satellites due to the selection criteria of the isolated sample.
Moreover, isolated and non-isolated satellites within $\tide<10\re$ are shown in the $1^{th}-4^{th}$ rows of the upper sub-table and in the $19^{th}-22^{th}$ rows of the lower sub-table, respectively. There are only two satellites with $\ar<1.0$ in the upper sub-table and one in the lower sub-table. Interactions from the halo centres or nearby massive galaxies are responsible for the lopsidedness of the satellites close to the halo centres or massive galaxies. However, the LRA is not found even in these strongly tidally disrupted satellites. The reason could be the rotation of the satellite galaxies, similar to the case of NGC 891 \citep{Mapelli+2008}. For a satellite spinning approximately one-half cycle around its minor axis, the longer side of the major axis appears on the opposite side of the massive nearby galaxy or the host halo centre. Considering most of the satellites are oblate systems or discs (see Fig. \ref{theta}), the rotation of these satellites might be the reason for the absence of LRA. A systematic study on the rotation and angular momentum of satellites will be performed in a follow-up project.

At the bottom of the two sub-tables of Table \ref{tab:table_3}, there are nineteen isolated and six non-isolated satellites extremely lopsided and with $\tide> 10\re$. Tidal stripping from halo centres of nearby massive galaxies is not the driving force for the lopsidedness of these satellites. 

\subsection{Other mechanisms to produce lopsided satellites}\label{origins}

The external perturbations and internal dynamics accounting for the lopsidedness of galaxies have been extensively studied \citep[e.g.,][]{Bournaud+2005b,Mapelli+2008,Kenney+2015,Zaritsky+2013,Levine_Sparke1998,Noordermeer+2001,Weinberg1994,Jog1997}. We briefly discuss the possible origins of lopsidedness below.

\begin{figure}
  \includegraphics[width=90mm]{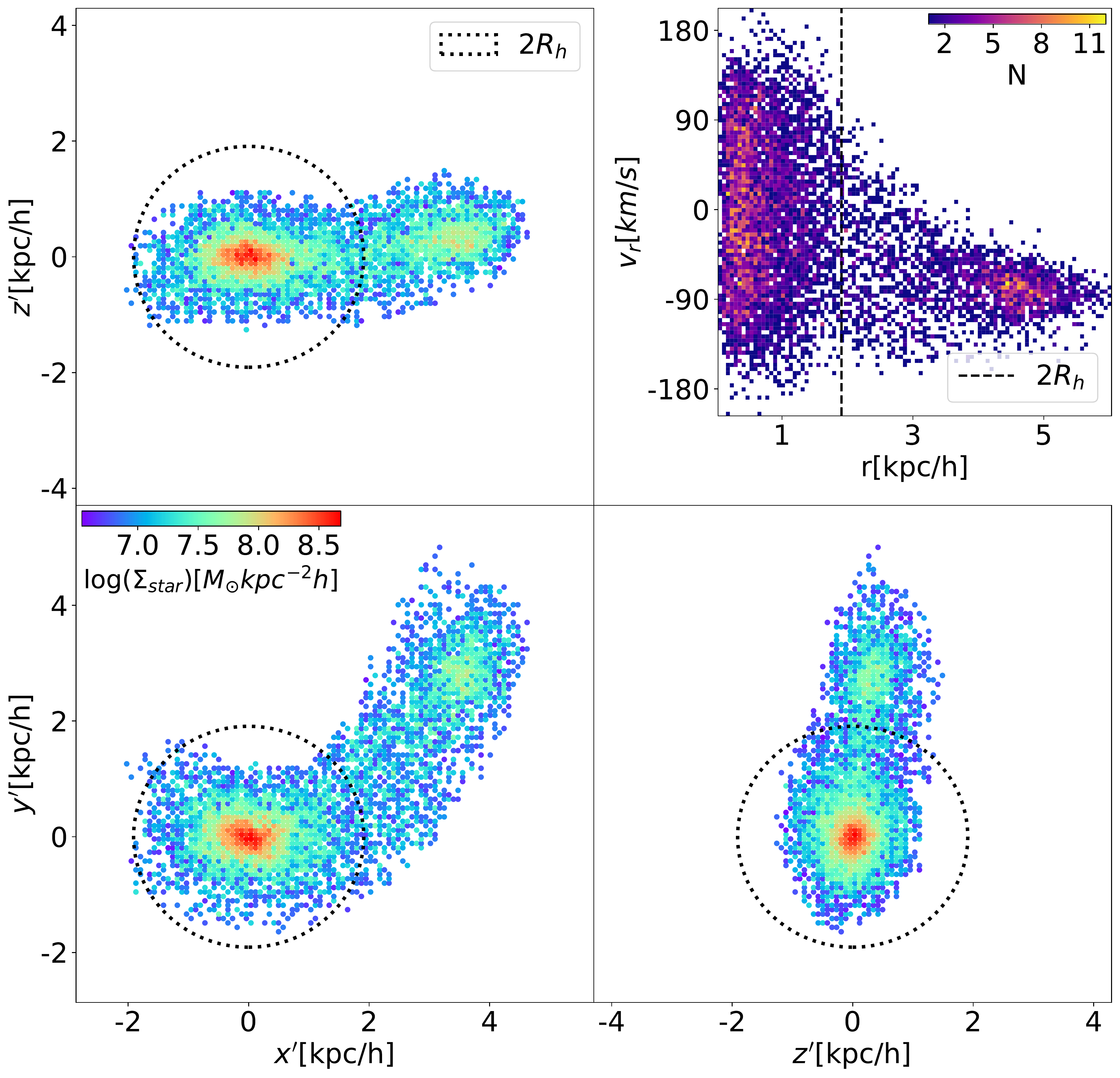}
  \caption{The projected stellar mass densities of satellite subID 566663 are exhibited on $x'-z'$ (upper left), $x'-y'$ (lower left) and $z'-y'$ (lower right) planes of inertia tensor's eigenframe, with black dotted circles showing the radii of $2\re$. This satellite's radial phase-space density distribution, $r$ versus $v_r$, is displayed in the upper right panel.}
  \label{merger}
\end{figure}

\subsubsection{External origins}
As aforementioned, the interactions from the halo centres or massive galaxies make a small part of satellites extremely lopsided, as summarised in Table \ref{tab:table_3}. These satellites reside in the central regions of their host haloes or the vicinity of nearby massive galaxies. The tidal radii of these galaxies are smaller than $2\re$. Moreover, there are a large number of satellites in Table \ref{tab:table_3} that do not have a dark matter subhalo.

Several dark-matter-free (DM-free) galaxies have been recently observed by \citet{Oh+2015,vanDokkum+2018a,vanDokkum+2018b,vanDokkum+2019}. 
The existence of DM-free galaxies is at odds with the standard stellar-to-halo mass relation, which challenges the standard $\Lambda$CDM framework \citep{Haslbauer+2019}. On the other hand, some research groups claimed that the DM-free galaxies are the remnants of a dwarf galaxy undergone tidal disruptions through close encounters \citep[e.g.,][]{Jing+2019,Moreno+2022}. Since the dark matter particles are dynamically hotter than the stellar particles in a dwarf galaxy, the dwarf galaxy loses its dark matter halo by the tidal shocks as the consequence of a close encounter with another galaxy \citep{Jing+2019}. But some other researchers found that dark matter-deficient galaxies form through encounters of disc galaxies \citep{Barnes_Hernquist1992}. These DM-free galaxies, formed in the tidal arms of the disc galaxies, are named tidal dwarf galaxies. We have examined the formation scenario of the DM-free galaxies appearing in Table \ref{tab:table_3}.

Below, we shall show four of the extremely lopsided DM-free satellites, including subID 300911, subID 435762 subID 535781 and subID 566663 in Table \ref{tab:table_3}. Such satellites are passing through or have just passed through the disc planes of their central host galaxies. The dark matter subhaloes of these satellites have been tidally disrupted by the central galaxy or a nearby massive galaxy. Although the dark matter subhalos of these galaxies seem to be removed by tidal shocking, more careful tests need to be performed by tracing their mass loss history backwards in time.

 We have examined the origins of the four example DM-free satellites. The galaxy subID 566663 forms through a merger between subID 563017 and subID 565398 at z=0.02 (the $97_{th}$ snapshot). At the redshift of z=0.02, subID 563017 is the more massive galaxy, whose progenitor, subID 554374, can be traced backwards in time until a redshift of z=0.08 (the $93_{rd}$ snapshot). Both progenitors of the DM-free galaxy, subID 566663 at the redshift of zero, are dark-matter deficient galaxies at a redshift of 0.08. Furthermore, we traced the more massive progenitor, subID 554374 at z=0.08, backwards in time. We found that $65\%$ stars of this progenitor form in gas-rich environments during the $90_{th}-92_{nd}$ snapshots. Thus the DM-free satellite galaxy, subID566663 at the present day, forms in the dark-matter deficient tidal arms. The other three example galaxies have similar tidal formation histories. In the satellite subID 300911, $99\%$ of stars form between $z=0.14$ and $z=0.13$ (the $88_{th}$-$89_{th}$ snapshots) through mergers between two satellite galaxies (subIDs 268865 and 268866) and the central galaxy (subID 268862). The DM-free spiral galaxies subIDs 535781 and 435762 formed recently at $z=0.01$ and $z=0.02$ (the $98_{th}$ and $97_{th}$ snapshots, respectively). The dark matter subhalos of these example galaxies are not disrupted by external interactions from nearby galaxies. Instead, they are tidally formed dwarf galaxies. These galaxies were born without dark matter subhalos.

The lopsidedness of the DM-free satellite galaxies may result from multiple origins. For instance, in the central regions of a cluster, satellites are perturbed by tidal stripping caused by the central galaxy. Meanwhile, the satellites are interacting with each other because the number density of galaxies is larger in the cluster centre. Considering their relative velocities are also higher when they are in a cluster's central regions, the gravitational interaction integration time is also shorter. Thus is it hard to draw a simple conclusion on the satellite-to-satellite-interaction-induced asymmetric shape. Let us take a DM-free satellite subID 566663 as an example galaxy. The distance between this satellite and its halo centre is $0.14\rvir$. The halo-centric position of the satellite is $(x', ~y', ~z')=(13.0, ~11.5, ~0.6)\kpc /h$ and the relative velocity to the halo centre is $(v_{\rm x'},~v_{\rm y'},~v_{\rm z'})=(-107.3, ~133.4, ~-8.6)\kms$, when the whole halo is rotated to the eigenframe of inertia tensor of the central galaxy. The orbit of this satellite on the scale of the cluster of galaxies implies that it is passing through the stellar plane of its host halo. The ratio $(\ar)_{2\re}=1.17~>1.0$. Moreover, we do not find any tidal tails in this satellite galaxy.
Therefore, the long semi-major axis is on the far side of its central galaxy, indicating that the lopsidedness is not due to tidal stripping. Other physical origins are required. To find out the true mechanism, we examine the stellar surface density of this satellite in the left two panels as well as the lower-right one of Fig. \ref{merger}. There is a clear structure located outside $2\re$. Moreover, an infalling clump at the same radius is found in the phase-space density distribution of subID 566663 displayed in the upper right panel of Fig. \ref{merger}. Since the particles within the clump are identified as part of the satellite subID 566663, the clump is infalling shows that the satellite is a non-fully relaxed merger remnant.

Early studies have demonstrated that asymmetric gas accretion with subsequent star formation is responsible for the lopsidedness of galaxies \citep[e.g.,][]{Bournaud+2005b,Mapelli+2008}. The cold gas accreted from the cosmic filaments could form new stars and thus lead to an asymmetric shape of the stellar component in a satellite galaxy \citep{Keres+2005}. We have examined the possible origins of the strongly lopsided galaxies in Table \ref{tab:table_3} and found that none of their lopsidedness is caused by star formation from gas accretion.

\begin{figure*}
  \includegraphics[width=160mm]{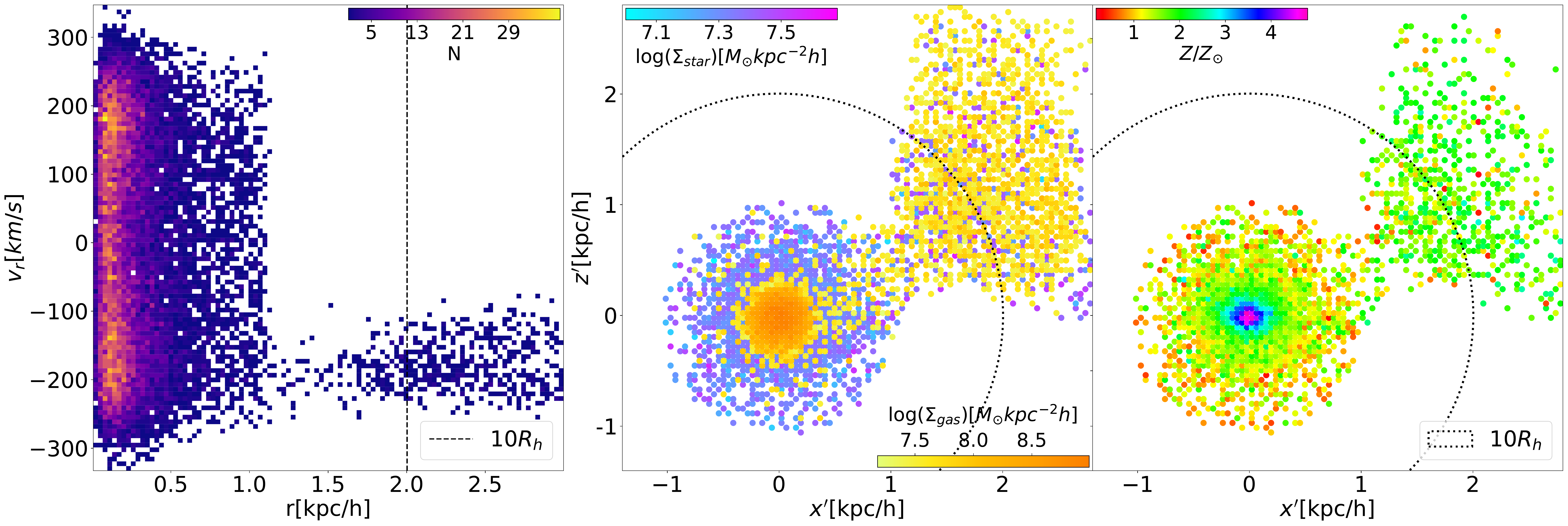}
  \caption{The phase-space density distribution of the satellite subID 455296 is displayed in the left panel. In the middle panel, the projected surface densities of the stellar and gaseous components are presented with colours provided in the two colour bars. The metallicity of the stellar component of the satellite is demonstrated in the right panel.}
    \label{gasaccretion}
\end{figure*}

On the other hand, the accretion of stars from external structures might contribute to the asymmetric of galaxies. A typical example satellite galaxy, subID 455296, is presented in Fig. \ref{gasaccretion}. The phase-space density distribution (left panel) reveals that there is a substructure falling into the satellite centre. The gas distribution in this satellite is examined (middle panel) to exclude other physical processes such as a merger of galaxies. On the scale of $\approx 1\kpc$, the projected stellar density appears to be quite symmetric on the $x'-z'$ plane of its eigenframe of inertia tensor. The semi-axial of the major axis is $1.02$ within $0$-$2\re$. Extended structures are in the upper right corner of the projected densities of stellar and gaseous components. The value of $A_1$ within the radial range of $1$-$2\re$ is $0.12$. The gas and stars are accreted into the satellite centre. The metallicity of stars in this satellite is studied in the right panel. The metallicity of stars in the regions of infalling gas appears to be lower, which suggests that the asymmetric comes from the accretion of stars from external structures.

\begin{figure}
  \includegraphics[width=\columnwidth]{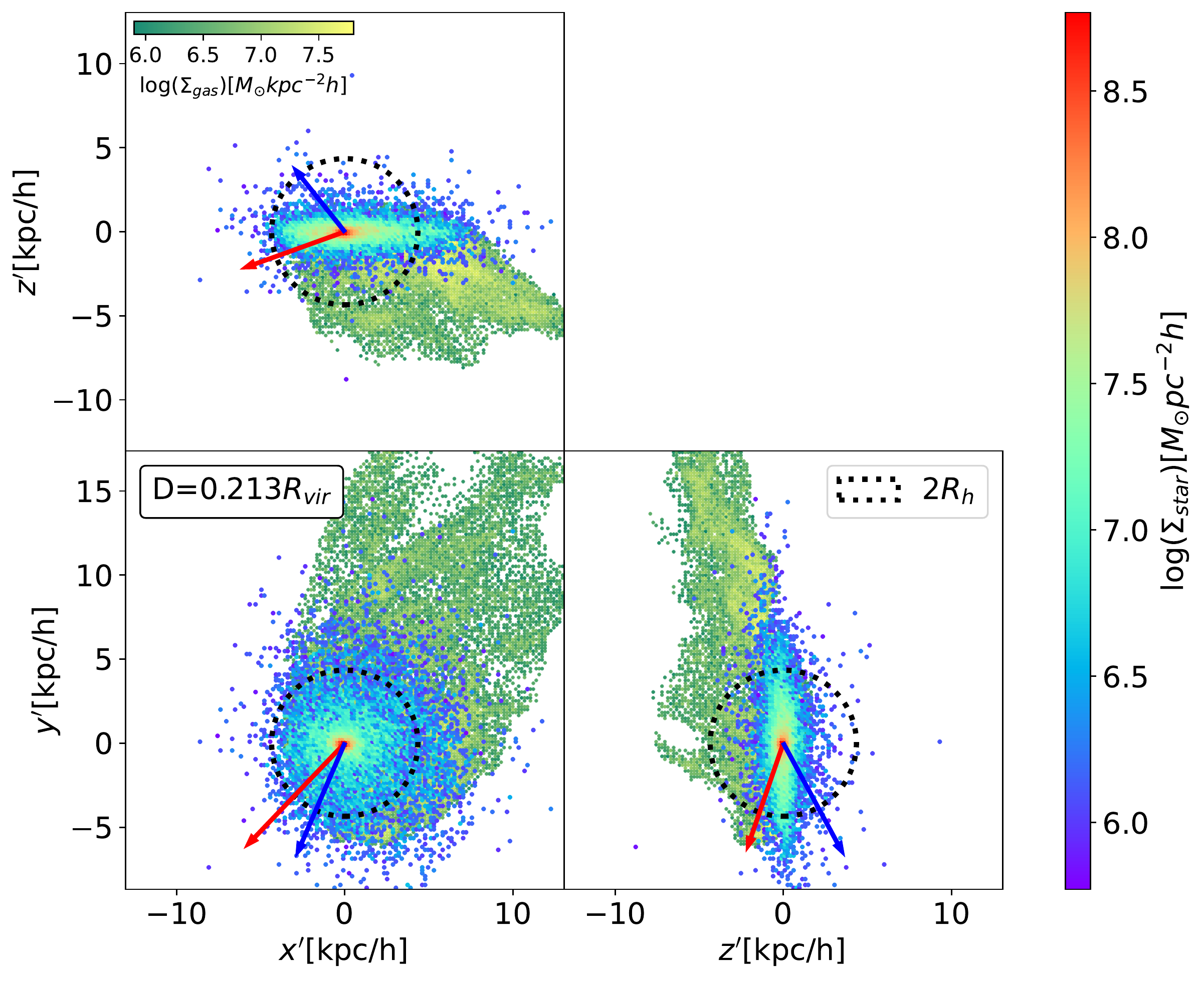}
  \caption{The projected stellar mass density (rainbow colours) and projected gaseous density (green-to-yellow colours) of the satellite subID 313700. An example of ram pressure stripping on a satellite galaxy. The two arrows indicate the directions of the halo centre (red arrow) and the motion of the satellite (blue arrow). The distance between the satellite and the halo centre is $0.21\rvir$. The black dashed circles show the position of $2\re$.}
    \label{ram}
\end{figure}

Ram pressure stripping from the diffuse intergalactic medium (IGM) plays only a minor role in producing the lopsidedness of galaxies \citep{Mapelli+2008}. The gaseous component in a satellite is compressed by the IGM in the leading side and is stretched in the opposite direction of the orbit \citep[e.g., a recent review by][]{Boselli+2022}. The gaseous component can be stripped outside the stellar component by ram pressure \citep{Smith+2012}. As a consequence, the stellar and dark matter components would be influenced by the drag force imposed by the gas, which in turn leads to an offset of a few $\kpc$ between the disc and the dark matter halo \citep{Smith+2012}. The satellite subID 313700 displayed in Fig. \ref{ram} is a representative example of a ram pressure-stripped lopsided galaxy. In the eigenframe of the inertia tensor of the host central galaxy, the galactocentric distance and relative speed between subID 313700 and the central galaxy are $74\kpc /h$ and $765.3\kms h^{-1}$, respectively. This satellite is in an incoming orbit to the central galaxy. The gaseous component of the satellite has been stripped out and left behind the stellar disc. Given the large gas fraction ($f_{\rm gas} =82\%$) in this satellite, the gravitational force introduced by the gaseous component is significant. The gravitation from the stripped gas then drags the stars in the direction opposite to the direction of motion. We, therefore, conclude that the lopsidedness of the stellar component is caused by ram pressure in the subID 313700-like satellites.

\subsubsection{Internal origins}

Apart from the external origins for the lopsidedness of galaxies, internal dynamics can also produce significantly lopsided galaxies. For instance, the gravitational instability of disc galaxies may result in a significantly lopsided shape of a galaxy \citep{Zaritsky+2013}. The asymmetric shapes of the strongly lopsided galaxies in Table \ref{tab:table_3} all have an external origin. Thus the internal perturbations only result in moderate or mild asymmetry for the satellite galaxies. For instance, the satellite subID 63869, is a typical isolated disc galaxy with spiral arms. The stellar disc is thin, with $c: a =0.34$ within the radial range of $0$-$2\re$. In addition, this satellite is far away from the central galaxy and other satellite galaxies. Moreover, there is neither asymmetric gas accretion nor recent minor merger events in this satellite. The semi-axial ratio of the major axis is $\ar =1.05$ within $0$-$2\re$. Thus the gravitational instability of the thin disc might be the mechanism of the lopsidedness.
Moreover, since this work mainly focuses on the alignments between the asymmetric shapes and the halo centres, the internal origins of lopsidedness will not be discussed further.

\subsection{Asymmetric tidal tails in observations and symmetric tidal tails in simulations}
Now we briefly review some recent progress on tidal tails which are complementary to our present work.
Observations on Palomar 5 streams showed that the tidal structures for this star cluster are asymmetry \citep{Bernard+2016}. Very recently, there are more asymmetric tidal tails observed in star clusters, including Hyades, Praesepe, Coma Berenices, COIN-Gaia 13 and NGC 752 \citep{Kroupa+2022}. 
However, the tidal tails of satellites are found to be extremely symmetric in a smooth and static host halo potential in previous studies of Newtonian dynamics \citep{Dehnen+2004}. The sole tidal field from the Milky Way cannot reproduce the asymmetric tidal tails of Palomar 5 \citep{Dehnen+2004}, unless other consequences of gravitational interactions are introduced, such as perturbations induced by giant molecular clouds \citep{Amorisco+2016}, subhaloes \citep{Erkal+2017}, spiral arms \citep{Banik_Bovy2019} or a prograde Galactic bar \citep{Pearson+2017}. \citet{Bonaca+2020} emphasised that none of the currently existing scenarios can explain all the observational features of the stellar streams of Palomar 5. To explain all the observational results of Palomar 5 within one scenario is out of the scope of this work. The symmetric tidal tails of satellite galaxies predicted in Newtonian dynamics agree with our results on the absence of the LRA.

\section{Summaries and conclusions}\label{conclusions}
In this work, we have presented a detailed analysis of the semi-major axial ratios ($\ar$) of the satellite galaxies generated by the TNG50-1 simulation. We have found that the distribution of $\ar$ is almost symmetric around the central value of $\ar=1.0$. This result indicates that the probabilities for the long semi-major axes of satellites to point towards and outwards the halo centres are nearly equal. We have adopted two criteria to define the lopsidedness of a satellite. More concretely, a satellite galaxy is regarded as lopsided if $\Delta >0.1$ or $A_1>0.2$. Under the former criterion, the fractions of strongly lopsided satellites are $7.5\%$ and $11\%$ for the isolated and non-isolated satellites truncated at $2\re$, respectively. Under the latter criterion, the two numbers become $7.8\%$ and $18.1\%$, respectively. While the fractions of strongly lopsided satellites in our analysis are at least one order of magnitude larger than those reported in \citet{Lokas2022}, they are still smaller than observations \citep[e.g.,][]{Jog_Combes2009,vanEymeren+2011,Yu+2022}.

The correlations between $\ar$ and other physical parameters have been further studied. We have shown that $\ar$ does not correlate with the distance to the halo centres, $D$, in both the inner and outer regions of the satellites. The fraction of galaxies with long semi-major axes in the near side to halo centres, $f_{N_-}$, is almost equal to $50\%$. These findings imply that the tidal fields from the halo centres do not lead to the lopsidedness of satellites. Moreover, we have found that neither the masses of the host haloes $M_{\rm halo}$ nor the satellite masses $M_{\rm sat}$ correlate with $\ar$. However, we have revealed a correlation between the values of $\ar$ and the fractions of gas, $f_{\rm gas}$, in the satellites. In both isolated and non-isolated samples of satellites, more massive galaxies tend to be more gas-poor. Moreover, the more gas-rich galaxies are more lopsided, with a larger $\Delta a$.

 In this work, we have explored the existence of RA in an isolated sample of satellite galaxies. The RA signal within the inner and intermediate regions of satellite galaxies is so weak that the existence of radial alignment cannot be asserted. However, RA is statistically significant present in the outer regions. Our findings are in agreement with the previous cosmological simulations \citep{Kuhlen+2007,Faltenbacher+2008,Pereira+2008,Knebe+2020} as well as certain observations \citep{Hawley_Peebles1975,Thompson1976,Pereira_Kuhn2005,Faltenbacher+2007a,Wang+2008,Wang+2019}.

We have known from the above analysis that the lopsidedness of satellites is not mainly induced by the tidal fields from the halo centres. Although this result provides us with evidence that LRA is unlikely to present in the simulated satellites within the framework of $\Lambda$CDM. More careful investigation is required to reach a conclusive answer. This motivated us to examine the existence of such a new alignment in the TNG50-1 simulation. The alignment angle, $\cos\theta$, defined in Eq. \ref{aa} has been systematically analysed. No LRA has been observed in all radial ranges.

Finally, we have briefly discussed the possible origins of the lopsidedness of satellites in the TNG50-1 simulation. The lopsidedness could be induced by the tidal stripping from the nearby central galaxies or massive satellite galaxies, the recent mergers, the asymmetric accretion of stellar streams from external structures and the ram pressure stripping. The internal origins such as gravitational instability only lead to moderate or mild lopsidedness in the TNG50-1 satellite galaxies.

\section{Acknowledgements}
The authors acknowledge the public release of Illustris-TNG data. We thank the anonymous referee and the statistics editor for the constructive and detailed comments. We appreciate Dr Hongsheng Zhao (St Andrews)'s suggestions regarding the alignment angle and the Roche lobes issues. We want to express our heartfelt gratitude to the contributors of the CRAN R project. Their work has provided powerful tools that have enabled us to understand better and interpret our data. We deeply respect and appreciate their hard work. Additionally, we would like to extend our thanks to the open-source packages, including {\it twosamples}, {\it cramer} \citep{Rizzo2019,Feigelson_Babu2012} and {\it circular} \citep{Fisher+1993,Mardia_Jupp2000,Pewsey+2013,Pewsey2018,Ley_verdebout2017,Mohammad2021}. We appreciate the efforts of all those involved in creating and maintaining these valuable resources.

XW is financially supported by the Natural Science Foundation of China (Number NSFC-12073026, NSFC-11421303) and ``the Fundamental Research Funds for the Central Universities''. YZ is financially supported by the ``Fund for Fostering Talents in Basic Science of the National Natural Science Foundation of China NO.J1310021''.

XW motivated and supervised the project. JS and XW analysed the released TNG50-1 data with the assistance of BG. YZ contributed to the early analysis of Illustris-1 data as an undergraduate thesis project. XW and JS wrote the manuscript with contributions from BG and YZ.

\bibliographystyle{aasjournal}
\bibliography{Reference}

\end{document}